\documentclass{ptephy}

\usepackage{boites}
\usepackage{graphicx}
\usepackage{wrapft}
\usepackage{wrapfig}

\newcommand{\bra}[1]{\langle {#1} |}
\newcommand{\ket}[1]{| {#1} \rangle}
\newcommand{\inproduct}[2]{\langle #1 | #2 \rangle}

\begin{document}

\title{
Density functional approaches to collective phenomena in nuclei:
Time-dependent density-functional theory for
perturbative and non-perturbative nuclear dynamics}

\author{Takashi Nakatsukasa}

\address{
RIKEN Nishina Center, Wako-shi 351-0198, Japan
}

\begin{abstract}%
We present the basic concepts and our recent developments
in the density functional approaches with the Skyrme functionals
for describing nuclear dynamics at low energy.
The time-dependent density-functional theory (TDDFT) is utilized for
the exact linear response with an external perturbation.
For description of collective dynamics beyond the perturbative regime,
we present a theory of a decoupled collective
submanifold to describe for a slow motion based on the TDDFT.
Selected applications are shown to demonstrate the
quality of their performance and feasibility.
Advantages and disadvantages in the numerical aspects
are also discussed.
\end{abstract}


\parindent0pt

\maketitle

\section{Introduction}

\subsection{Nucleus as a quantum object}

The nuclei provide the matters with mass,
the stars with fuel, and the universe with a variety of elements.
It was discovered by Ernest Rutherford and coworkers about 100 years ago
\cite{Rut1911}, that explains the large-angle scattering of alpha particles 
by a gold foil \cite{GM1909}.
This discovery was also a beginning of the era of the quantum mechanics.
Rutherford estimated the upper limit of the nuclear size
which turned out to be much smaller than that of the atom.
In the atomic scale (\AA), the nuclear size (fm)
could be regarded as just a point!
According to the classical mechanics,
the atoms must collapse into nuclear size, because
the attractive Coulomb potential eventually brings all the
electrons into the nucleus.
This mystery stimulated Niels Bohr to develop his idea on the quantum
mechanics \cite{Bohr1913}.

\subsubsection{Atoms and molecules}

With a knowledge of the quantum mechanics,
it is easy to understand why the atoms do not collapse to the nuclear scale.
If an electron was confined in the nuclear scale of femtometer,
the uncertainty principle immediately tells us that
its zero-point kinetic energy would become gigantic (order of GeV).
In order to decrease this kinetic energy to a reasonable magnitude,
the electron's wave function must have an atomic size of \AA.
Thus, the atomic size is a consequence of
the quantum effect.

The molecules (and the solids) are a good contrast to atoms.
The atom is bound by the Coulomb interaction,
whose range is infinite ($V_{N-e} \sim -Ze^2/r$),
between
a positively charged nucleus and electrons with a negative charge.
Since the molecules consist of these charge-neutral atoms,
the interaction between a pair of neutral atoms does not have
the long-range tail of $1/r$, but normally has
a short-range repulsive part and an intermediate-range attractive part
(Fig. \ref{fig:aa-nn}(a)).
Because of this charge neutrality,
the molecule is easy to disintegrate into smaller units
(molecules and atoms).
The atomic size is approximately constant and independent from the atomic
number, while the molecular size varies depending on the number of atoms
and their kinds.
Last but not the least, in the zero-th order approximation,
the ground states of the molecules can be
classically described as atoms located at fixed relative positions.
The hindered quantum fluctuation in molecules is simply due to
the fact that the atomic mass, which is approximately
identical to the nuclear mass, is about 2,000 times larger than the
electronic mass.

\subsubsection{Nuclei}

\begin{figure}[tb]
\centering
\includegraphics[width=0.8\textwidth]{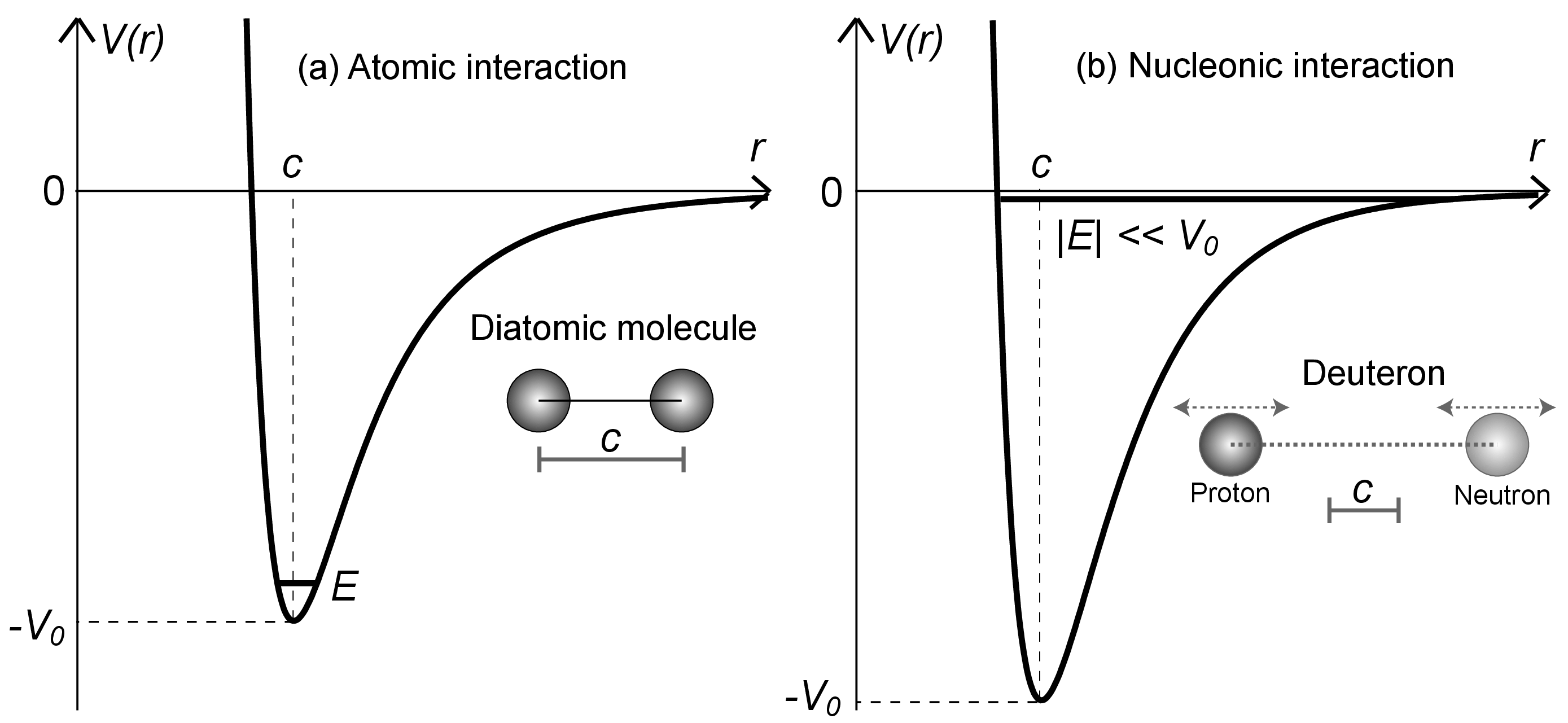}
\caption{
Schematic representation of atomic and nucleonic interactions $V(r)$
as a function of the relative distance $r$.
The energy of the bound (ground) state $E$ is shown by a horizontal line.
(a) Atom-atom interaction:
$V_0=1\sim 10$ eV and $c\approx a_B$, where $a_B$ is the Bohr radius.
(b) Nucleon-nucleon interaction:
$V_0\approx 100$ MeV and $c=0.5\sim 1$ fm.
The essence of the present figure is taken
from Fig. 2-36 in Ref. \cite{BM69}.
}
\label{fig:aa-nn}
\end{figure}

The nucleus has a number of properties analogous to the molecules,
except for its strong quantum nature.
It is a self-bound system composed of fermions of spin $1/2$ and
isospin $1/2$ with
approximately equal masses, called nucleons (protons and neutrons).
The nuclear species are classified by the numbers of neutrons ($N$)
and protons ($Z$).
Rutherford discovered that the size of the nucleus is as tiny as
the femtometer, but later it was found that the nuclear size varies,
as its volume is roughly proportional to the mass number ($A=N+Z$).
Each nucleon is a color-singlet (neutral) object.
The interaction between a pair of nucleons (nuclear force) has a
finite range of a scale of the pion Compton wave length ($\lambda_\pi$).
Similar to the molecules, it has 
a short-range repulsive part and an intermediate-range attractive part
(Fig. \ref{fig:aa-nn}(b)).
The nucleus can be disintegrated into small pieces
with a small separation energy.
In fact, heavy nuclei can have ``negative'' separation energies because
of the repulsive Coulomb energy among protons.

The quantum nature is an important difference between the nucleus and
the molecule.
In Fig.~\ref{fig:aa-nn}, we show schematic pictures of the atomic
and nucleonic potentials.
The ground state of the diatomic molecule (panel (a))
is formed at the bottom of the potential, $E\approx -V_0$.
In the length scale of \AA,
the atomic mass is heavy enough to localize the wave
function at the location of the bottom of the potential, $r\approx c$.
Thus, the relative distance between a pair of atoms is fixed at $r=c$.
This property allows us to describe the atomic motion in the
classical mechanics, such as in the molecular dynamics.
In contrast, the nuclear interaction is not strong enough to bind
nucleons at the bottom of the potential.
In other words, the nucleon's mass is too light to localize the
wave function in the sub-femtometer scale.
In deuteron, the zero-point kinetic energy cancels the
negative potential energy ($\langle T +V \rangle \approx 0$),
leading to a bound state at approximately
zero energy ($E=-2.2\mbox{ MeV}\gg-V_0$).
The deuteron wave function spatially extends beyond the range of
nuclear force ($\sim \lambda_\pi$),
which reduces the kinetic energy $\langle T \rangle$.
This shows a striking contrast to the diatomic molecule,
and is even analogous to the atoms, that
the large size of the deuteron is a consequence of the quantum effect.
This strong quantum nature also tells us that
the infinite nuclear matter will not be crystallized
even at zero temperature, but will stay as the liquid.
The nuclear system provides us with unique opportunities to study
femto-scale quantum liquids.

\subsection{Computing nucleus from scratch}

The strong quantum nature in finite nuclei leads to a rich variety
of unique phenomena.
Remarkable experimental progress in production and study of exotic nuclei
requires us to construct theoretical and computational approaches
with high accuracy and a reliable predictive power.
Extensive studies have been made for
constructing theoretical models to elucidate basic nuclear dynamics behind
a variety of nuclear phenomena.
Simultaneously, significant efforts have been made in the microscopic
foundation of those models.

For light nuclei,
the ``first-principles'' large-scale computation,
starting from the bare nucleon-nucleon (two-body \& three-body) forces,
is becoming a current trend in theoretical nuclear physics.
Among them,
the Green's function Monte Carlo (GFMC) method is
the most successful first-principles approach
to nuclear structure calculation \cite{PW01}.
In this approach, using the Monte Carlo technique,
the many-body wave function is sampled in the coordinate space with
spin and isospin degrees of freedom.
The success of the GFMC clearly demonstrates that we are able to construct
a light nucleus from the scratch on the computer.
The GFMC method has been applied to nuclei up to the mass number $A\approx 10$.
Another first-principles approach is to project the nuclear
Hamiltonian in a truncated Hilbert space, then diagonalize it.
This is called no-core shell-model (NCSM) method \cite{NNBV06}.
The NCSM also shows successful applications up to the $p$-shell nuclei.
The GFMC and NCSM both indicate the exponential increase in computational
tasks with respect to the increasing nucleon number.
The third approach, the coupled-cluster method (CCM), has an advantage
that the required computation
increases only in power with respect to the nucleon number.
The CCM, which was originally invented in nuclear physics \cite{KLZ78}
and later became extremely successful in quantum chemistry \cite{BM07},
has been revisited as an ab-initio computational approach
to nuclear structure \cite{Wlo05}.
Especially, it is powerful to study closed-shell nuclei.

Although these first-principles approaches
have recently shown a significant progress,
they are still limited to nuclei with the small mass number.
This may sound mysterious to physicists in other fields, because
we know that the similar kinds of approaches are able to treat systems of
much larger particle numbers.
For instance, using the CCM, nowadays,
the chemists can easily calculate a molecular structure with 100 electrons.
Why is the first-principles calculation of nuclear structure so difficult?
The answer is perhaps trivial for nuclear physicists, but may not be so for
others.
Let us pick up several important aspects leading to this answer.
(1) The strong quantum nature.
As we have discussed previously, the full quantum mechanical treatment
is necessary for nuclear structure calculation.
(2) Strong coupling nature.
The nucleon-nucleon scattering length is approximately $a\approx -18$ fm in
$^1S_0$ channel.
This is much larger than the mean distance between nucleons inside
the nucleus ($k_F|a| \gg 1$).
(3) Singular property of the nucleonic interaction.
To reproduce the phase shift for the nucleon-nucleon scattering,
the interaction should have a strong repulsive core at short distance
\cite{BM69}.
(4) Complexity of the nucleonic interaction.
The interaction has a strong state dependence which may be represented
by the spin- and isospin-dependence \cite{BM69}.
It also contains strong non-central and non-local terms,
such as tensor and spin-orbit interactions.
Furthermore, for a quantitative description of nuclei,
it is indispensable to introduce three-body interactions
in addition to the two-body force.
(5) Coexistence of different interactions.
In addition to the strong interaction, we need to treat the
electromagnetic interactions among protons.
(6) Coexistence of different energy scales.
The nuclear binding energy amounts to order of GeV for heavy nuclei.
Because of the strong quantum nature in nuclei,
this is a consequence of the cancellation between
positive kinetic energy and negative potential energy.
Thus, we need to compute these enormous positive and negative components
in high accuracy to obtain a correct binding energy.
(7) Fermionic nature of nucleons.
Needless to say, since the nucleons are fermions, 
the total wave function must be anti-symmetrized.
(8) Finite systems without an external potential.
Electronic many-body problems in molecules and solids are solved
with external fields produced by nuclei with positive charges.
In contrast, the nucleus is a self-bound finite system.
Thus, we are usually required to obtain an intrinsic wave function without
the center-of-mass degrees of freedom.
This requirement often restricts our choice of the basis functions.

Despite of these difficulties,
significant advances in the computer power may lead to
a realistic ``first-principles'' construction of the $sd$-shell nuclei
in near future.
There have been extensive efforts toward this direction,
especially developing an algorithm suitable for parallel use of
the vast number of processors \cite{unedf}.

\subsection{Density functional theory (DFT)}

In contrast to the first-principles calculations, which are
limited to nuclei with small number of nucleons,
the density functional theory (DFT) is currently
a leading theory for describing
nuclear properties of heavy nuclei \cite{BHR03,LPT03}.
It is capable of
describing almost all nuclei, including nuclear matter,
with a single universal energy density functional (EDF).
An argument based on the quantum many-body theory leading to nuclear EDF
was also developed in 70's$-$80's \cite{Neg82,Nak11}.
In addition, its strict theoretical foundation is given by
the basic theorem of the DFT \cite{HK64,KS65}.
Since the nucleus is a self-bound system without an external
potential, we should slightly modify the DFT theorem.
This will be discussed in Sec. \ref{sec: basic_formalism}.

The nucleus by itself produces a potential confining nucleons
which is analogous to the Kohn-Sham potential in DFT \cite{KS65}.
Nuclear physicists often call this potential ``mean field'',
though it is different from a naive mean-field potential
directly constructed from the nucleonic interaction.
There are many evidences for the fact that the mean-free
path of nucleons inside the nucleus is significantly
larger than the nuclear radius \cite{BM69},
in spite of the strong (and even singular) two-body interaction.
This is partially due to the Pauli exclusion principle.
Since the nucleon mean-free path is roughly proportional to
$\left\{ \epsilon_F / (\epsilon-\epsilon_F) \right\}^2$ \cite{Gal58},
it is significantly enhanced for nucleons whose
energies are close to the Fermi energy.
Another even simpler argument was given by Bohr and Mottelson \cite{BM69},
that the nuclear normal density is much lower
than the one giving the close packing limit (crystalline limit).
In this argument, the quantum effect plays a primary role.

The DFT theorem guarantees the existence of
generalized density functionals for every physical observable
(see Sec.~\ref{sec: basic_formalism}).
However, to construct the exact functional,
we need experimental data and other theoretical inputs.
Currently, there are a variety of EDFs
which predict somewhat different properties for nuclei very far
from the stability line.
We are in want of finding an ultimate universal energy density
functional, which is capable of exact description of
every nucleus in the nuclear chart.
In addition to the recent progress in the first-principles
calculations for light nuclei,
the radioactive beam facilities in the world will give us
ideal opportunities to determine the parameter set for a better
functional.
New data on neutron halos and skins in medium heavy nuclei
may provide important information
on its dependence on density and density gradient.
Observation of new isotopes in a long isotopic (isotonic) chain may
lead to useful constraints on the isovector parts of the energy functional.

An extension of the DFT to the time-dependent DFT (TDDFT) provides
a feasible description of many-body dynamics, which contains information
on excited states in addition to the ground state.
The TDDFT is justified by the one-to-one correspondence between the
time-dependent density and time-dependent external potential \cite{RG84},
which will be presented in Sec.~\ref{sec: TDDFT}.
The TDDFT has vast applications to quantum phenomena in many-body systems.
Among them, the perturbative regime has been mostly studied so far.
Different approaches to the linear response calculations will be presented
in Sec.~\ref{sec: perturbative_regime}.
It is of significant interest but challenging to go beyond the
perturbative regime.
Nuclei show numerous phenomena related to the large amplitude collective
motion, such as fission, shape transition, shape coexistence,
anharmonic vibrations, and so on.
We present, in Sec.~\ref{sec: LACM},
 a theory to identify an optimal collective submanifold
in the classical phase space of large dimensions.

\section{Basic formalism for particles in self-bound systems}
\label{sec: basic_formalism}

The density functional theory (DFT) has been extremely successful
for calculations of ground-state
properties of atoms, molecules, and solids.
It describes a many-particle system
exactly in terms of its one-body density alone.
The DFT is based on the original theorem of Hohenberg and Kohn (HK) \cite{HK64}
which was proved for the ground-state of the many-particle system.
Every observable can be written, in principle, as
a functional of density.

In nuclear physics, however, we need to treat an isolated system
without an external potential.
The present nuclear EDF
produces a localized density profile without an external potential.
Thus, the ground state spontaneously
violates the translational symmetry and seems to contain
spurious excitation related to the center-of-mass motion.
Furthermore, it also violates the rotational symmetry when the
ground state is spontaneously deformed.
It is interesting to see whether the nuclear EDF
can be theoretically justified in a strict sense.
We present a possible justification in Sec.~\ref{sec: DFT_wave_packet},
according to the recent progress \cite{Eng07,Gir08,GJB08}.

\subsection{DFT theorem for a wave-packet state}
\label{sec: DFT_wave_packet}

The HK theorem \cite{HK64} guarantees one-to-one mapping between a one-body
density $\rho(\vec{r})$ and an external potential $v_0(\vec{r})$.
Then, since the ground-state wave function is a functional of density,
in principle,
all the observables should be functionals of the density as well.
However, to describe an isolated self-bound finite system
in a box of volume $V$,
it is somewhat useless to use the ground-state density
in the laboratory frame, because it must be constant,
$\rho(\vec{r})=\rho=N/V\rightarrow 0$ ($V\rightarrow\infty$),
where $N$ is the particle number.
Instead, we want to use a functional of the {\it intrinsic} density
$\rho(\vec{r}-\vec{R})$, where $\vec{R}$ is
the center-of-mass coordinate of the total system.
In this case, the original HK theorem cannot be justified, because
it adopts a one-body external field $v_0(\vec{r})$ coupled to
the density $\rho(\vec{r})$ in the laboratory frame.

Validity of the DFT for the intrinsic states has been discussed by several
authors \cite{Eng07,Gir08,GJB08}.
In this section, we present a method to define the DFT for an intrinsic state,
more precisely for a ``wave-packet'' state.
The argument here essentially follows the idea by
Giraud et al \cite{GJB08}.

In fact,
all the nuclear EDFs, currently available,
produce a {\it wave-packet} state.
The minimization of an EDF $E[\rho]$
without the external potential ($v_0(\vec{r})\rightarrow 0$)
leads to the nucleon density distribution $\rho(\vec{r})$ with
a finite radius.
This violates the translational symmetry of the original Hamiltonian.
Therefore, it is of our interest here to justify
the energy functional of
the {\it wave-packet} density in the laboratory frame
in a strict sense.

First, we assume that a wave-packet state in the laboratory frame
can be expressed by a product wave function of
{\it intrinsic} and {\it spurious} degrees of freedom,
$\ket{\Phi}=\ket{\phi}\otimes\ket{\chi}$.
Here, $\ket{\phi}$ indicates the intrinsic state
and $\ket{\chi}$ defines the spurious motion.
This decomposition can be exactly done for the translational motion.
\begin{eqnarray}
\label{H_original}
&& H = H_{\rm intr}(\xi,\pi) + \frac{\vec{P}^2}{2Nm}, \\
&&\Phi(\vec{r}_1, \cdots, \vec{r}_N)
= \phi(\vec{\xi}_1, \cdots, \vec{\xi}_{N-1}) \otimes \chi (\vec{R}) ,
\end{eqnarray}
where $\vec{R}$ $(\vec{P})$ denote the center-of-mass coordinates
(total momenta).
$\vec\xi_i$ and $\vec{\pi}_i$ are the relative Jacobi coordinates
and their conjugate momenta, respectively.
Since the intrinsic ground state, which is supposed to be unique,
is completely independent from the spurious motion,
we can adopt an arbitrary form of $\chi(\vec{R})$;
e.g., $\chi(\vec{R})\propto \exp(-R^2/2b^2)$.
Then, the ground wave-packet state
can be obtained by the variation after the projection:
\begin{equation}
\label{chi-fixed}
(E_0)_\chi \equiv
\min_{\chi{\rm :fixed}}
\left[ \frac{\bra{\Phi}H P\ket{\Phi}}{\bra{\Phi}P\ket{\Phi}}
\right]
\end{equation}
where the projection operator $P$ does not change the intrinsic state
$\ket{\phi}$ but makes $\chi(\vec{R})$ an eigenstate of the total linear
momentum $P=0$.
The variation with respect to the full space
($\ket{\Phi}=\ket{\phi}\otimes\ket{\chi}$) contradicts
the uniqueness of the ground state,
because states with different $\chi(\vec{R})$ give
the same $P\ket{\Phi}$.
Therefore, the variation here should be performed with respect only to
the intrinsic state $\ket{\phi}$.
This is indicated by the subscript ``$\chi$:fixed'' in Eq. (\ref{chi-fixed}).
The wave-packet density profile is simply given by
$
\rho(\vec{r})\equiv \bra{\Phi} \hat{\psi}^\dagger(\vec{r})
                                   \hat{\psi}(\vec{r})\ket{\Phi}
$,
that depends on the form of $\chi(\vec{R})$.
In the followings, we always assume 
a fixed form of $\chi(\vec{R})$ for
the wave-packet state $\ket{\Phi}$.

Next, we introduce an external potential $V_0=\sum_i v_0(\vec{r}_i)$.
The following minimization with respect to the intrinsic state $\ket{\phi}$
leads to the ``minimum'' energy $E_0[v_0]$
and defines the wave packet $\ket{\Phi}$.
\begin{equation}
E_0[v_0]=
\min_{\chi{\rm :fixed}}
\left[ \frac{\bra{\Phi}H P\ket{\Phi}}{\bra{\Phi}P\ket{\Phi}}
+\bra{\Phi} V_0 \ket{\Phi}
\right]
=\min_{\chi{\rm :fixed}}
\left[ \frac{\bra{\Phi}H P\ket{\Phi}}{\bra{\Phi}P\ket{\Phi}}
+\int \rho(\vec{r})v_0(\vec{r}) d^3r 
\right] .
\end{equation}
Noted that $V_0$ operates on a state $\ket{\Phi}$,
not on a projected state $P\ket{\Phi}$.
$E_0[v_0]$ does not correspond to the ground-state
energy of a system with the Hamiltonian $H+V_0$,
however, it reduces to Eq. (\ref{chi-fixed}) for $V_0=0$.
Now, let us show the one-to-one correspondence between the external
potential $V_0$ and the wave-packet density $\rho(\vec{r})$.
The proof proceeds by {\it reductio ad absurdum} in the same manner
as the original proof of HK \cite{HK64}.
Assume that another potential $v'_0(\vec{r})$, which defines
the wave packet $\ket{\Phi'}$, produces the same density
$\rho(\vec{r})$.
Then, the energy for $\ket{\Phi'}$ is given by
\begin{equation}
E_0[v_0']=
\frac{\bra{\Phi'}H P\ket{\Phi'}}{\bra{\Phi'}P\ket{\Phi'}}
+\bra{\Phi'} V_0' \ket{\Phi'} .
\end{equation}
If we replace the state $\ket{\Phi'}$ by $\ket{\Phi}$,
the energy must increase.
\begin{eqnarray}
E_0[v_0']
&<&
\frac{\bra{\Phi}H P\ket{\Phi}}{\bra{\Phi}P\ket{\Phi}}
+\bra{\Phi} V_0' \ket{\Phi} \nonumber \\
&=&
\frac{\bra{\Phi}H P\ket{\Phi}}{\bra{\Phi}P\ket{\Phi}}
+\bra{\Phi} V_0 \ket{\Phi}
+\bra{\Phi} V_0'-V_0 \ket{\Phi}
\nonumber\\
&=&
E_0[v_0] +\int \rho(\vec{r})\left\{v_0'(\vec{r})-v_0(\vec{r})\right\} d^3r .
\label{E'<E}
\end{eqnarray}
Interchanging $V_0$ and $V_0'$, we also find
\begin{equation}
\label{E<E'}
E_0[v_0] <
E_0[v_0'] +\int \rho(\vec{r})\left\{v_0(\vec{r})-v_0'(\vec{r})\right\} d^3r .
\end{equation}
Addition of Eqs. (\ref{E'<E}) and (\ref{E<E'}) leads to the
inconsistency, $E_0[v_0]+E_0[v_0'] < E_0[v_0] + E_0[v_0']$.
This proves the one-to-one correspondence between the external field
$V_0$ and the wave-packet density $\rho(\vec{r})$.
Thus, both $v_0(\vec{r})$ and the wave packet $\ket{\Phi}$
are functionals of the density $\rho(\vec{r})$.
In order to lift restriction to $v$-representative densities,
we use the constrained search \cite{Lev79} in which one considers
only states that produce a given density $\rho(\vec{r})$,
and define the universal functional
\begin{equation}
\label{F_rho}
F_\chi[\rho]\equiv
\min_{\Phi\rightarrow\rho}
\left[ \frac{\bra{\Phi}H P\ket{\Phi}}{\bra{\Phi}P\ket{\Phi}}
\right] .
\end{equation}
Here, the subscript ``$\Phi\rightarrow\rho $'' indicates the
minimization with a constraint on $\rho(\vec{r})$.
The density functional $F_\chi[\rho]$ contains
the energy of the fixed center-of-mass spurious motion, $E_\chi$,
that is trivially given by
$
E_\chi(N) \equiv \bra{\chi} \vec{P}^2\ket{\chi}/(2Nm) 
$.
Therefore, the energy of the ground state with $P=0$
(intrinsic energy) may be obtained by
the minimization with a constraint on the total particle number, as
\begin{equation}
E_{\rm intr}(N) \equiv E_{P=0}(N) =
\min_{\rho} \left[ F_\chi[\rho]
- \lambda \left( \int \rho(\vec{r}) d^3r - N \right)\right] 
- E_\chi (N)  .
\end{equation}
Note that, although the wave function $\ket{\chi}$ is fixed,
$E_\chi$
may depend on the total mass of the system (particle number $N$).
In principle, the expectation value of any observable $\hat{O}$,
which only depends on
the intrinsic degrees of freedom, is a functional of $\rho$,
\begin{equation}
{\mathcal O}[\rho] \equiv
\bra{\phi} \hat{O} \ket{\phi} \times \inproduct{\chi}{\chi}
=\bra{\Phi} \hat{O} \ket{\Phi} ,
\end{equation}
because the wave-packet state $\ket{\Phi}$ is given as
a functional of $\rho$.
Since the state $\ket{\chi}$ is fixed, there is a trivial correspondence
between the wave-packet density $\rho(\vec{r})$ and
the intrinsic density $\rho(\vec{r}-\vec{R})$.
This completes the basic theorem of the DFT for the wave packet.

\subsection{The Kohn-Sham (KS) scheme}

For a many-body system of fermions, the shell effects
play a major role to determine the ground state.
In other words, we need a density functional which takes account of
the kinetic energy properly.
This is known to be difficult in the local density approximation \cite{RS80}.
At present, a scheme given by Kohn and Sham \cite{KS65}
only provides a practical solution for this problem.
Here, we follow the same argument.

We introduce a {\it reference system} which is a
``virtual'' non-interacting system with
an external potential $v_s(\vec{r})$.
This reference system is supposed to reproduce
the same density $\rho(\vec{r})$ of
the ``physical'' interacting wave packet, but does not have to reproduce
the center-of-mass wave function $\ket{\chi}$.
The ground state of the reference system is trivially obtained as
a Slater determinant constructed by the solution of\footnote{
Precisely speaking, the orbitals $\phi_i$ are
not necessarily the eigensolutions of Eq. (\ref{KS_eq}), but
arbitrary as far as they give the same Slater determinant.
We come back to this gauge freedom in Sec.~\ref{sec: TDKS}.
}
\begin{equation}
\label{KS_eq}
\left(-\frac{1}{2m}\nabla^2 + v_s(\vec{r}) \right)
 \phi_i (\vec{r})= \epsilon_i \phi_i(\vec{r}) ,
\end{equation}
adopting the unit $\hbar=1$,
and the density is given by
$\rho(\vec{r})=\sum_{i=1}^N |\phi_i(\vec{r})|^2$.
The kinetic energy\footnote{
The HK theorem guarantees that
$T_s[\rho]$ of the reference system is
a functional of the density.
}
is given by
\begin{equation}
\label{KS_kinetic}
T_s[\rho]=\sum_{i=1}^N \bra{\phi_i} \left(-\frac{1}{2m} \nabla^2 \right)
\ket{\phi_i} .
\end{equation}
The variation of the total energy of the reference system
\begin{eqnarray}
\label{KS_energy}
E_s[\rho] = T_s[\rho] + \int v_s(\vec{r})\rho(\vec{r}) d^3r ,
\end{eqnarray}
with a constraint on the particle number,
$\delta(E_s[\rho]-\mu \int \rho(\vec{r})d^3r)=0$, leads to the
following equation:
\begin{eqnarray}
\label{Euler_eq}
\mu = \frac{\delta T_s[\rho]}{\delta\rho(\vec{r})} + v_s(\vec{r}) .
\end{eqnarray}
Although we did not explicitly construct $T_s[\rho]$ as a functional
of $\rho(\vec{r})$,
the solution of Eq.~(\ref{Euler_eq}) must be identical to the solution
of Eqs. (\ref{KS_eq}) and (\ref{KS_kinetic}).

The success of the Kohn-Sham (KS) scheme comes from a simple idea to
decompose the kinetic energy in the physical interacting system 
into two parts;
$T_s[\rho]$, which is a major origin of the shell effects,
and the rest, which is treated as a part of ``correlation energy''
described by a simple functional of density,
\begin{eqnarray}
F_\chi[\rho] = T_s[\rho] + E_c[\rho] ,
\end{eqnarray}
where $E_c[\rho]\equiv F_\chi[\rho] - T_s[\rho]$.
Then, the variation of $F_\chi[\rho]$ leads to Eq.~(\ref{Euler_eq})
but the potential is now a functional of density, defined by
$v_s(\vec{r})\equiv \delta E_c[\rho] / \delta\rho(\vec{r})$.
The only practical difference between the reference system and
the interacting system
is that, since $v_s(\vec{r})$ is a functional of density in the latter,
Eq. (\ref{KS_eq}) must be self-consistently solved.
These equations are called Kohn-Sham (KS) equations.
The self-consistent solution of the KS equations
provides the density $\rho(\vec{r})$ of a wave-packet state
with a fixed $\ket{\chi}$ corresponding to a (local) minimum of
the EDF, $F_\chi[\rho]$.
The success of the KS scheme is attributed to the goodness of
the local density approximation for $E_c[\rho]$.

\subsection{Open issues}

\subsubsection{Subtraction of the center-of-mass energy}

In the proof of the basic theorem for the wave packet, given in
Sec.~\ref{sec: DFT_wave_packet},
we need to fix a wave function of the center-of-mass motion $\ket{\chi}$.
The choice of this spurious wave function is arbitrary, but
the energy $E_\chi$ depends on this choice.
In practice, the subtraction of $E_\chi$ is normally performed
by constructing the state $\ket{\chi}$ from the obtained result.
This is somewhat inconsistent with the assumption of
the fixed center-of-mass state $\ket{\chi}$.
This could be easily corrected by taking $E_\chi$ of a given $\ket{\chi}$.
However, as far as we know, this has not been examined yet.

\subsubsection{Validity of the Kohn-Sham scheme}

The KS scheme is to take into account a major part
of the kinetic energy as $T_s[\rho]$, and the rest as a correction.
In other words, the KS scheme implicitly assumes that the energy
$E_c[\rho]$ is able to be well approximated by a simple functional of $\rho$.
In fact, this question is still an open issue, not only
in the nuclear physics
but also in other quantum many-body systems.
In the present wave-packet theory,
the kinetic energy of the wave packet,
$T_\chi[\rho]=\bra{\Phi}\hat{T}\ket{\Phi}$,
depends on the center-of-mass state $\chi(\vec{R})$.
Therefore, there may be an optimal choice for $\chi(\vec{R})$
to minimize the difference $T_\chi[\rho]-T_s[\rho]$.
The question about the validity of the Kohn-Sham scheme
remains to be answered.

\subsubsection{Non-spherical wave packet}

The nuclear EDFs are known to produce
a spontaneous symmetry breaking about the rotational symmetry.
Namely, we often encounter a deformed wave-packet density,
which accounts for appearance of the rotational spectra in nuclei.
For instance, many experimental evidences indicate that
nuclei in the rare-earth region and in the actinide region are deformed \cite{BM75}.
According to the argument in Sec.~\ref{sec: DFT_wave_packet},
we may separate the rotational motion from the intrinsic degrees
of freedom, then, we have
\begin{equation}
\Phi(\vec{r}_1,\cdots,\vec{r}_N)\approx \phi(\xi)
 \otimes \chi(\vec\Theta,\vec{R}) ,
\end{equation}
where $\vec\Theta$ indicates angle variables.
Then, replacing the operator $P$ by that of angular momentum projection,
the DFT for deformed wave packet can be shown in the same manner.
However, in contrast to the translational motion,
the separation of the rotation degrees of freedom is not exact.
but only approximate.
Thus, there remains an ambiguity for the definition of the functional
(\ref{F_rho}):
Namely, the minimization must be performed in the entire space except for
the subspace that accounts for translational and rotational correlations.
In this sense, the use of the EDF, which produces a deformed state,
can be justified only approximately.

\subsection{Time-dependent density functional theory}
\label{sec: TDDFT}

The DFT is designed for calculating the ground-state properties.
For excited-state properties and reactions,
the time-dependent density functional theory (TDDFT)
is a powerful and useful tool.
In this section, we recapitulate the basic theorem for
the time-dependent density functional theory (TDDFT).

Since the proof of the HK theorem is based on the
Rayleigh-Ritz variational principle,
its extension to the time-dependent density is  not straightforward.
This was done by Runge and Gross \cite{RG84},
showing
that there is one-to-one correspondence between a time-dependent density
$\rho(\vec{r},t)$ and a time-dependent external potential $v(\vec{r},t)$.
The external potential is required to be expandable in a Taylor series about
the initial time $t_0$,
\begin{equation}
v(\vec{r},t)=\sum_{k=0}^\infty  \frac{1}{k!} v_k(\vec{r}) (t-t_0)^k .
\end{equation}
The external potentials, $v(\vec{r},t)$ and $v'(\vec{r},t)$,
 are defined to be {\it different} if
there exist some minimal nonnegative integer $k$ such that
$\nabla w_k(\vec{r})\neq 0$ where
$w_k(\vec{r})\equiv v_k(\vec{r})-v'_k(\vec{r})$.
In other words, $v(\vec{r},t)$ and $v'(\vec{r},t)$
differ more than a time-dependent function,
$v(\vec{r},t)-v'(\vec{r},t)\neq c(t)$.
Note that the potentials differing by the time-dependent constant $c(t)$
produce the same density $\rho(\vec{r},t)$ because the corresponding
wave functions differ by a merely time-dependent phase,
as in Eq.~(\ref{Psi_t}) with $\alpha(t)=\int^t c(s) ds$.

Now, let us assume that starting a common initial state
$\ket{\Psi_0}=\ket{\Psi(t_0)}$,
two different external potentials, $v(\vec{r},t)$ and $v'(\vec{r},t)$,
produce densities $\rho(\vec{r},t)$ and $\rho'(\vec{r},t)$,
respectively.
From this, we first prove that the current densities,
$\vec{j}(\vec{r},t)$ and $\vec{j'}(\vec{r},t)$,
are different.
Using the current density operator $\vec{\hat{j}}(\vec{r})$,
the equation of motion is written as
\begin{equation}
\label{dj_dt}
i\frac{\partial}{\partial t} \vec{j}(\vec{r},t)
= \bra{\Psi(t)} [ \vec{\hat{j}}(\vec{r}), \hat{H}(t) ] \ket{\Psi(t)} ,
\end{equation}
where $\hat{H}(t)=H_0+\sum_{i=1}^N v(\vec{r}_i,t)$.
We have the same equation for $\vec{j'}(\vec{r},t)$,
with $\ket{\Psi(t)}$ and $v(\vec{r},t)$ replaced by
$\ket{\Psi'(t)}$ and $v'(\vec{r},t)$, respectively.
Then, we have
\begin{eqnarray}
\left. \frac{\partial}{\partial t}
\left\{\vec{j}(\vec{r},t)-\vec{j'}(\vec{r},t)\right\} \right|_{t=t_0}
&=& -i \bra{\Psi_0} [ \vec{\hat{j}}(\vec{r}),
                    \hat{H}(t_0)-\hat{H}'(t_0) ] \ket{\Psi_0}
  \nonumber\\
&=& - \frac{1}{m}\rho(\vec{r},t_0) \nabla w_0(\vec{r}) .
\end{eqnarray}
If $\nabla w_0(\vec{r})\neq 0$, it is easy to see that
$\vec{j}(\vec{r},t)$ and $\vec{j'}(\vec{r},t)$ are different at $t>t_0$.
In case that $\nabla w_0(\vec{r})= 0$ and $\nabla w_1(\vec{r})\neq 0$,
we need to further calculate derivative of Eq. (\ref{dj_dt})
with respect to $t$.
\begin{equation}
\label{d2j_dt2}
\left. i\frac{\partial^2}{\partial t^2} \vec{j}(\vec{r},t) \right|_{t=t_0}
= \bra{\Psi_0} [ \vec{\hat{j}}(\vec{r}),
        \frac{\partial \hat{H}(t_0)}{\partial t} ] \ket{\Psi_0}
+ \bra{\Psi_0} [ [ \vec{\hat{j}}(\vec{r}), \hat{H}(t_0) ], \hat{H}(t_0) ]
  \ket{\Psi_0} .
\end{equation}
The second term of Eq. (\ref{d2j_dt2}) vanishes for
$\partial^2/\partial t^2\{\vec{j}(\vec{r},t) - \vec{j'}(\vec{r},t)\}|_{t=t_0}$,
because $\hat{H}'(t_0)=\hat{H}(t_0)+\mbox{const}$.
Thus,
\begin{equation}
\left. \frac{\partial^2}{\partial t^2}
\left\{\vec{j}(\vec{r},t)-\vec{j'}(\vec{r},t)\right\} \right|_{t=t_0}
= - \frac{1}{m}\rho(\vec{r},t_0) \nabla w_1(\vec{r}) \neq 0.
\end{equation}
Again, we can conclude that $\vec{j}(\vec{r},t)\neq \vec{j'}(\vec{r},t)$
at $t>t_0$.
In general, if $\nabla w_k(\vec{r})= 0$ for $k<n$ and
$\nabla w_n(\vec{r})\neq 0$, we repeat the same argument to reach
\begin{equation}
\label{dn1j_dtn1}
\left. \left(\frac{\partial}{\partial t}\right)^{n+1}
\left\{\vec{j}(\vec{r},t)-\vec{j'}(\vec{r},t)\right\} \right|_{t=t_0}
= - \frac{1}{m}\rho(\vec{r},t_0) \nabla w_n(\vec{r}) \neq 0.
\end{equation}
Therefore, there exists a mapping from the expandable potential $v(\vec{r},t)$
to the current density $\vec{j}(\vec{r},t)$.

Next, we use the continuity equation
\begin{equation}
\label{continuity}
\frac{\partial}{\partial t} \left\{\rho(\vec{r},t)-\rho'(\vec{r},t)\right\}
=-\nabla \cdot \left\{ \vec{j}(\vec{r},t)-\vec{j'}(\vec{r},t) \right\} ,
\end{equation}
and calculate the (n+1)-th derivative of Eq. (\ref{continuity})
at $t=t_0$.
From Eq. (\ref{dn1j_dtn1}),
\begin{equation}
\label{rho_n2}
\left.
\left(\frac{\partial}{\partial t}\right)^{n+2}
\left\{\rho(\vec{r},t)-\rho'(\vec{r},t)\right\}
\right|_{t=t_0}
=\frac{1}{m}\nabla \cdot \left\{ \rho(\vec{r},t_0) \nabla w_n(\vec{r})\right\} .
\end{equation}
Provided $\nabla w_n(\vec{r})\neq 0$, we can prove that
the right-hand side of Eq. (\ref{rho_n2}) does not vanish identically.
This is done by using the following identity:
\begin{equation}
\label{identity}
\nabla \cdot \left\{ 
       \rho(\vec{r},t_0) w_n(\vec{r}) \nabla w_n(\vec{r}) \right\}
-w_n(\vec{r}) \nabla \cdot\left\{ \rho(\vec{r},t_0) \nabla w_n(\vec{r})
\right\}
= \rho(\vec{r},t_0)\left\{ \nabla w_n(\vec{r}) \right\}^2  .
\end{equation}
The integral of both sides of Eq. (\ref{identity}) over the entire space leads
to
\begin{equation}
-\int d\vec{r}
 w_n(\vec{r}) \nabla \cdot\left\{ \rho(\vec{r},t_0) \nabla w_n(\vec{r})
\right\}
=  \int d\vec{r}
\rho(\vec{r},t_0)\left\{ \nabla w_n(\vec{r}) \right\}^2  > 0 ,
\end{equation}
where we assume $\rho(\vec{r},t_0)$ is localized in space so that
the surface integral vanishes.
Therefore, from Eq. (\ref{rho_n2}), we can conclude that the densities
$\rho(\vec{r},t)$ and $\rho'(\vec{r},t)$ are different at $t>t_0$.
This completes the proof.

The time-dependent density determines the time-dependent external potential
except for the time-dependent constant.
Therefore,
the many-body time-dependent state should be a functional of density
except for a time-dependent phase.
\begin{equation}
\label{Psi_t}
\ket{\Psi(t)}=\exp(-i\alpha(t)) \ket{\Psi[\rho](t)} .
\end{equation}
Any observable quantity must be independent from the global phase, $\alpha(t)$,
thus, a unique functional of density, $A[\rho(t)]$.
Note that these time-dependent density functionals
depend on the initial many-body state $\ket{\Psi_0}$.

\subsection{Time-dependent Kohn-Sham (TDKS) equations}
\label{sec: TDKS}

In practice, we use the Kohn-Sham scheme \cite{KS65} for numerical calculations.
Assuming the $v$-representability,
the time-dependent Kohn-Sham (TDKS) equation is given by
\begin{equation}
\label{TDKS}
i\frac{\partial}{\partial t}\psi_i(\vec{r},t) = \left\{
-\frac{1}{2m}\nabla^2 + v_s[\rho](\vec{r},t) \right\}
\psi_i(\vec{r},t) ,
\quad i=1,\cdots,N .
\end{equation}
The density of a system is expressed by
$\rho(\vec{r},t) = \sum_{i=1}^N \left| \psi_i(\vec{r},t)\right|^2$.
In practice, we usually adopt the potential same as the
one for calculation of the ground state (``adiabatic approximation''),
except for the external potential $v(\vec{r},t)$.
\begin{equation}
\label{TD_v_s}
v_s[\rho](\vec{r},t)=v(\vec{r},t)
+\left.\frac{\delta E[\rho]}{\delta\rho(\vec{r})}\right|_{\rho}
\end{equation}

The density
is invariant with respect to the
unitary transformation $U(N)$ among $N$ occupied KS orbitals.
Therefore, there are gauge degrees of freedom to choose this
transformation at any instant of time.
For explicit notification of the gauge freedom,
it is convenient to introduce the matrix notation as follows.
Let $\{ \ket{\alpha} \}$ be an arbitrary single-particle basis set,
and we define a matrix $\bar\psi$ of size of $\infty\times N$, as
$\bar\psi_{\alpha i}(t)\equiv \inproduct{\alpha}{\psi_i(t)}$.
Then, the density matrix $\rho(t)$ can be written as
\begin{equation}
\label{rho_psi}
\rho_{\alpha\beta}(t)=\sum_i \inproduct{\alpha}{\psi_i(t)}\inproduct{\psi_i(t)}{\beta}
= \left( \bar\psi(t) \bar\psi^\dagger(t) \right)_{\alpha\beta} .
\end{equation}
The orthonormal property of the KS orbitals is expressed as
$\bar\psi^\dagger(t) \bar\psi(t) = 1$.
Denoting the TDKS Hamiltonian
in Eq. (\ref{TDKS}) as
$h_s(t)=h_s[\rho(t)]=-{\nabla^2}/(2m) + v_s[\rho](\vec{r},t)$,
the TDKS equations (\ref{TDKS}) can be generalized into the following form.
\begin{equation}
\label{G-TDKS}
i\frac{\partial}{\partial t} \bar\psi(t) = h_s(t) \bar\psi(t) - \bar\psi(t)\xi(t) ,
\end{equation}
where $\xi(t)$ is an arbitrary $N\times N$ time-dependent Hermitian matrix
which represents a generator of the $U(N)$ transformation.
Equation~(\ref{G-TDKS}) is equivalent to
the well-known equation for the density matrix  \cite{Eba10}.
\begin{equation}
\label{TDKS_density}
i\frac{\partial}{\partial t} \rho(t) =
\left[ h_s(t), \rho(t) \right] .
\end{equation}
The stationary state corresponds to the time-indenpendent density,
$\partial\rho/\partial t = 0$.

\subsection{Pairing correlations: Kohn-Sham-Bogoliubov (KSB) equations}

The HK theorem (or its wave-packet version), in principle, guarantees that
the energy of the system can be exactly written as a functional of density,
$E[\rho]$.
However, in practice, it is often difficult to take into account
all the correlation energy in $E[\rho]$, solely depending on $\rho(\vec{r})$.
The kinetic energy is such an example, which is resolved by
the genius idea by Kohn and Sham.
The pairing correlation energy $E_{\rm pair}$,
which is important for heavy nuclei
in open-shell configurations,
is another example difficult to be expressed by $\rho(\vec{r})$ only.

To overcome this difficulty, a common strategy is to extend the
KS equations, according to the Bogoliubov's quasiparticles \cite{Bog58}.
Each orbital now has two components,
$\Phi_\nu=\begin{pmatrix} U_\nu \\ V_\nu\end{pmatrix}$,
and its number is basically infinite ($\nu=1,\cdots,\infty$).
These are called quasiparticle (qp) orbitals.
The previous KS equations are extended to the following equations,
which we call Kohn-Sham-Bogoliubov (KSB) equations\footnote{
Again, this corresponds to a special choice in the gauge degrees of freedom.
See Sec.~\ref{sec: TDKSB}.
} hereafter:
\begin{equation}
\label{KSB_equation}
\left( {\mathcal H}_s - \mu {\mathcal N} \right)
\Phi_\nu = E_\nu \Phi_\nu ,
\end{equation}
where
\begin{equation}
\label{KSB_Hamiltonian}
{\mathcal H}_s \equiv
\begin{pmatrix}
h[\rho,\kappa] & \Delta[\rho,\kappa] \\
-\Delta^*[\rho,\kappa] & -h^*[\rho,\kappa]
\end{pmatrix}
, \quad\quad
{\mathcal N}=
\begin{pmatrix}
1 & 0 \\
0 & -1
\end{pmatrix}
.
\end{equation}
The Hamiltonian $h$ is in the same form as that in Eq. (\ref{KS_eq}),
$h=-\nabla^2/(2m) + v_s[\rho,\kappa]$,
while the pair potential $\Delta[\rho,\kappa]$ is introduced to
describe the pairing correlations.
The same form of equation is known as the Hartree-Fock-Bogoliubov equation
in nuclear physics \cite{RS80,BR86}.
Now, the KSB Hamiltonian in Eq. (\ref{KSB_Hamiltonian}) not only depends
on the density $\rho$, but also on the pair density $\kappa$.
In Eq.~(\ref{KSB_equation}), the matrix convention is assumed.
Namely, when we adopt a single-particle basis of $\{ \ket{\alpha} \}$,
we have $U_{\alpha\nu}\equiv\inproduct{\alpha}{U_\nu}$
and $V_{\alpha\nu}\equiv\inproduct{\alpha}{V_\nu}$, and
$h_{\alpha\beta}$ and
$\Delta_{\alpha\beta}$ correspond to Hermitian and anti-symmetric
matrices, respectively.

The chemical potential $\mu$ is determined so as to
satisfy the following number condition:
$\mbox{tr}\rho=N$.
With this number constraint,
$\mbox{tr}\rho$ must have a finite value.
On the other hand, the values of the pair density $\kappa$ are
determined solely by the variation of the total energy.
Therefore, the self-consistent solution of the KSB equations
(\ref{KSB_equation})
spontaneously produces the finite values of $\kappa$ and $\Delta$.

The solutions of the KSB equation have a ``paired'' property:
If the qp state
$\Phi_\nu=\begin{pmatrix} U_\nu \\ V_\nu\end{pmatrix}$
is a solution of Eq. (\ref{KSB_equation}) with eigenvalue $E_\nu$,
the qp state
$\bar\Phi_\nu=\begin{pmatrix} V_\nu^* \\ U_\nu^*\end{pmatrix}$
is also a solution with eigenvalue $-E_\nu$.
We call $\Phi_\nu$ ``unoccupied'' qp orbitals,
and $\bar\Phi_\nu$ ``occupied'' qp orbitals \cite{DFT84}.
This naming is based on the generalized density matrix,
\begin{equation}
R=
\begin{pmatrix}
\rho & \kappa \\
-\kappa^* & 1-\rho^*
\end{pmatrix}
\end{equation}
which is Hermitian and idempotent: $R^2=R$.
The ``unoccupied'' (``occupied'') qp orbitals correspond to the
eigenvectors of $R$ with eigenvalue 0 (1);
$R\Phi_\nu=0$ and $R\bar\Phi_\nu = \bar\Phi_\nu$.

Denoting the dimension of the single-particle Hilbert space as $M$,
we may define the $2M\times M$ matrix $\Phi$ as follows.
\begin{equation}
\label{Phi_matrix}
\Phi_{\alpha\nu}(t)=\begin{cases}
\inproduct{\alpha}{U_\nu} & \alpha=1,\cdots,M \\
\inproduct{\alpha-M}{V_\nu} & \alpha=M+1,\cdots,2M
\end{cases}
\end{equation}
which represents ``unoccupied'' qp orbitals
($\nu=1,\cdots,M$).
The ``occupied'' orbitals $\bar\Phi$ with size of $2M\times M$
are defined in the same manner,
with $U_\nu$ ($V_\nu$) replaced by $V_\nu^*$ ($U_\nu^*$).
The generalized densities are expressed in terms of these matrices as
$R=\bar\Phi\bar\Phi^\dagger=1-\Phi\Phi^\dagger$.
The orthonormal property of the qp orbitals is given by
$\Phi^\dagger \Phi = \bar\Phi^\dagger\bar\Phi=1$.
Combining the `unoccupied'' and ``occupied'' orbitals to construct
the $2M\times 2M$ matrix ${\mathcal W}\equiv (\Phi,\bar\Phi)$,
the matrix ${\mathcal W}$ becomes a unitary matrix  \cite{RS80}.

In the energy functional of Skyrme type \cite{VB72},
 the pairing correlation energy is
simply added to the original energy functional.
\begin{equation}
E[\rho,\kappa] \equiv E[\rho] + E_{\rm pair}[\rho,\kappa] ,
\end{equation}
that depends only on the local densities.
Therefore, Eq.~(\ref{KSB_equation}) becomes local in coordinates.
However, in general, the KSB Hamiltonian, $h$ and $\Delta$,
are not necessarily local.
For instance, the Gogny functional \cite{DG80} gives non-local KSB equations.

\subsection{Time-dependent Kohn-Sham-Bogoliubov (TDKSB) equations}
\label{sec: TDKSB}

For a time-dependent description,
the inclusion of the pair density
leads to the time-dependent Kohn-Sham-Bogoliubov (TDKSB) equations.
They can be formulated in a matrix form analogous to Eq. (\ref{G-TDKS}).
Using an arbitrary $2M\times M$ Hermitian matrix $\Xi(t)$,
the TDKSB equations may be written as
\begin{equation}
\label{G-TDKSB}
i\frac{\partial}{\partial t} \Psi(t) = {\mathcal H}_s(t) \Psi(t) - \Psi(t)\Xi(t) ,
\end{equation}
where the TDKSB Hamiltonian is given by Eq. (\ref{KSB_Hamiltonian}),
which depends on time through the densities $\rho(t)$ and $\kappa(t)$.
Here,
$\Psi(t)$ represent time-dependent ``unoccupied'' qp orbitals
($\nu=1,\cdots,M$).
The ``occupied'' orbitals $\bar\Psi(t)$
are defined in the same manner,
with $U_\nu$ ($V_\nu$) replaced by $V_\nu^*$ ($U_\nu^*$).
The TDKSB equation (\ref{G-TDKSB}) holds for $\bar\Psi(t)$, as well.

Analogous to the stationary case,
the generalized density $R(t)$ can be written as
$R(t)=\bar\Psi(t)\bar\Psi^\dagger(t) =
1-\Psi(t)\Psi^\dagger(t)$,
and the orthonormal property of the qp orbitals is given by
the unitarity of the $2M\times 2M$ matrix ${\mathcal W}(t)$.
The ``unoccupied'' $\Psi(t)$
(``occupied'' $\bar\Psi(t)$) correspond to the subspace with eigenvalue 0 (1);
$R(t)\Psi(t)=0$ and $R(t)\bar\Psi(t)=\bar\Psi(t)$.
In the generalized density matrix formalism,
the TDKSB equation is written in an analogous manner to
Eq. (\ref{TDKS_density}):
\begin{equation}
\label{TDKSB_density}
i\frac{\partial}{\partial t} R(t) =
\left[ {\mathcal H}_s(t), R(t) \right] .
\end{equation}

So far, we have shown similarities between
Eqs. (\ref{G-TDKS})$-$(\ref{TDKS_density}) and
(\ref{G-TDKSB})$-$(\ref{TDKSB_density}).
However, there is an important difference between Eq. (\ref{TDKS_density})
and Eq. (\ref{TDKSB_density}).
The stationary solution in Eq. (\ref{TDKS_density})
corresponds to $\partial\rho/\partial t=0$.
In contrast, it is not the case in Eq. (\ref{TDKSB_density}),
$\partial R/\partial t \neq 0$.
Let us examine this difference in details.
The TDKSB equation (\ref{G-TDKSB}) can be recast into another
form, convenient for taking its stationary limit.
First, let us factor out the time-dependent phases as follows:
$\Psi(t) = \exp(-i\mu{\mathcal N}t) \Psi'(t)$ and
$\bar\Psi(t) = \exp(-i\mu{\mathcal N}t) \bar\Psi'(t)$.
Here and hereafter,
we denote the remaining parts of the quantities as
the ``primed'' ones.
The generalized density becomes
\begin{equation}
R(t)=\bar\Psi(t)\bar\Psi^\dagger(t)
=\exp(-i\mu{\mathcal N}t) R'(t) 
\exp(+i\mu{\mathcal N}t)
,
\end{equation}
where
\begin{equation}
R'(t)=\bar\Psi'(t)\bar\Psi^{'\dagger}(t)
=\begin{pmatrix}
\rho(t) & \kappa'(t) \\
\kappa^{'*}(t) & 1-\rho^*(t)
\end{pmatrix}
.
\end{equation}
Namely, the transformation $\Psi(t)\rightarrow \Psi'(t)$
does not change the density $\rho$, but modifies the pair density as
$\kappa(t)=e^{-2i\mu t}\kappa'(t)$.
Since the pair potential $\Delta(t)$ is usually a linear functional of
$\kappa(t)$,
the same time-dependent phase should be assumed for $\Delta(t)$ as well:
$\Delta(t)=\Delta'(t) e^{-2i\mu t}$.
The Hamiltonian is transformed in the same manner:
\begin{equation}
{\mathcal H}_s(t) = 
\exp(-i\mu{\mathcal N}t) {\mathcal H}'_s(t) 
\exp(+i\mu{\mathcal N}t) .
\end{equation}
With these primed quantities, the TDKSB equation (\ref{G-TDKSB}) can be
rewritten as
\begin{equation}
\label{G-TDKSB-2}
i\frac{\partial}{\partial t} \Psi'(t) =
 \left\{{\mathcal H}'_s(t) - \mu {\mathcal N} \right\} \Psi'(t)
 - \Psi'(t)\Xi(t) ,
\end{equation}
or equivalently, in the generalized density matrix,
\begin{equation}
\label{TDKSB_density-2}
i\frac{\partial}{\partial t} R'(t) =
\left[ {\mathcal H}_s'(t)-\mu {\mathcal N}, R'(t) \right] .
\end{equation}

It is now clear that the stationary solution corresponds to
$\partial R'/\partial t=0$, not to $\partial R/\partial t=0$,
with a proper choice for the parameter $\mu$ identical to the
chemical potential.
In Eq. (\ref{G-TDKSB-2}), it corresponds to $\partial \Psi'/\partial t = 0$
with a choice of the $M\times M$ gauge matrix
$\Xi_{\nu\nu'}\equiv \bra{\Psi'_\nu} {\mathcal H}'_s - \mu{\mathcal N}
 \ket{\Psi'_{\nu'}}$.
It should be noted that the generalized density $R(t)$ is invariant
with respect to the choice of the gauge matrix $\Xi(t)$.
In contrast, the time-dependent phase factor in $\kappa(t)$ and $\Delta(t)$
have a physical origin and cannot be removed by the gauge choice.
In fact, it is a boost transformation, $e^{-i\mu {\mathcal N}t}$, from
the laboratory frame of reference to the body-fixed frame.
The stationary solution with $\kappa\neq 0$ ($\Delta\neq 0$)
corresponds to a time-dependent solution in the laboratory frame:
\begin{equation}
\Psi_\nu(t)=
\begin{pmatrix}
e^{-i\mu t} & 0\\
0 & e^{+i\mu t}
\end{pmatrix}
\Psi_\nu' 
 .
\end{equation}
This is a collective motion associated with the spontaneous generation
of the pair density, called pair rotation.
Therefore, in terms of the TDKSB formalism, the appearance of the
chemical potential in the stationary KSB equation (\ref{KSB_equation})
comes from the boost transformation to the body-fixed frame rotating
in the gauge space.
This is analogous to the appearance of the cranking term $-\omega J_x$
in the spatially rotating frame of reference \cite{RS80}.
In the case of pair rotation, since the particle number is finite
$N>0$, the system is rotating in the gauge space, even
at the ground state.
This rotation affects the {\it intrinsic} modes of excitation,
thus, the Hamiltonian in the rotating frame, ${\mathcal H}_s - \mu {\mathcal N}$,
should be utilized to calculate the {\it intrinsic} excitation spectra.
This point will be discussed again in Sec.~\ref{sec: MFHE}.

\section{Perturbative regime: Linear response}
\label{sec: perturbative_regime}

The theorem of the TDDFT tells us that the functional may depend
on the initial state, $\ket{\Psi(t_0)}$.
This additional ambiguity can be removed by assuming that the initial
state is identical to the ground state.
With this assumption, the linear response theory with a weak
time-dependent perturbation is formulated in this section.
The formulation is basically identical to the one known as
the random-phase approximation in nuclear physics \cite{RS80,BR86}.
However, according to the concept of the TDDFT,
the theory gives the exact linear density response,
with no approximation involved, in principle.\footnote{
In practice, some approximations are involved, such
as the adiabatic approximation of Eq. (\ref{TD_v_s}).
}

\subsection{Time-dependent linear density response}
We consider a system subject to a time-dependent external potential 
\begin{equation}
\label{v_ext}
v(\vec{r},t)=
\begin{cases}
0 & t < 0 \\
v_1(\vec{r},t) & t \geq 0 
\end{cases}
\end{equation}
in addition to the static potential $v_0(\vec{r})$ of the unperturbed
system.\footnote{
For isolated nuclear systems, we have $v_0=0$.
}
In this section, we use the notation of the four vector $x=(\vec{r},t)$.
We assume that the system is at the ground state at times $t< 0$.
Thus, the initial density $\rho_0(\vec{r})$ at $t\leq 0$ can be
obtained from the self-consistent solution
of the Kohn-Sham equations (\ref{KS_eq}).
The first-order density response, $\rho(x)\approx\rho_0(\vec{r})+\rho_1(x)$,
is given by
\begin{equation}
\label{rho_1}
\rho_1(x)=\int d^4x' \Pi(x,x') v_1(x')
\end{equation}
with the density-density response function
\begin{equation}
\label{Pi}
\Pi(x,x') = \left.
  \frac{\delta \rho(x)}{\delta v(x')} \right|_{v=0} .
\end{equation}
The right-hand side of Eq. (\ref{Pi}) is a well-defined quantity, since
the basic theorem of TDDFT in Sec.~\ref{sec: TDDFT} guarantees that
the time-dependent density is a functional of the time-dependent
external potential;
$\rho[v](x)$.

For non-interacting particles moving in an external potential
of $v_s(x)$, there is a one-to-one correspondence between
the time-dependent density and the potential.
Therefore, we have
\begin{equation}
\rho(x)=\rho[v_s](x), \quad\quad
v_s(x)=v_s[\rho](x) .
\end{equation}
The density-density response function for the non-interacting system is
given by
\begin{equation}
\label{Pi_s}
\Pi_s(x,x') = \left.
  \frac{\delta \rho(x)}{\delta v_s(x')} \right|_{v_s[\rho_0]} .
\end{equation}
The potential $v_s(x)$ is written as a sum of a given
external potential and the rest of the Kohn-Sham potential,
$v_s(x)=v(x)+v_{\rm ks}[\rho](x)$.
For instance, 
in the adiabatic approximation of Eq. (\ref{TD_v_s}),
$v_{\rm ks}(x)=\delta E[\rho]/\delta\rho(x)$.
Therefore, using the chain rules, Eq. (\ref{Pi}) can be connected to
its non-interacting $\Pi_s(x,x')$:
\begin{eqnarray}
\Pi(x,x') &=& \int d^4y \left.
  \frac{\delta \rho(x)}{\delta v_s(y)}
      \right|_{v_s[\rho_0]} \cdot
  \left.
  \frac{\delta v_s(y)}{\delta v(x')}
  \right|_{v=0} \nonumber \\ 
&=& \int d^4y 
\Pi_s(x,y) \left\{
\delta(y-x')
 + \int d^4y'
   \left.\frac{\delta v_{\rm ks}(y)}{\delta \rho(y')}\right|_{\rho_0}
   \left.\frac{\delta \rho(y')}{\delta v(x')}
  \right|_{v=0} \right\}
\nonumber \\ 
\label{Dyson_type_eq}
&=&
\Pi_s(x,x') +
\int d^4y \int d^4y'
\Pi_s(x,y) w(y,y') \Pi(y',x') ,
\end{eqnarray}
where the residual kernel is given by
\begin{equation}
w(x,x') \equiv \left.
  \frac{\delta v_{\rm ks}(x)}{\delta \rho(x')} \right|_{\rho_0} .
\end{equation}
In the adiabatic approximation, this is equal to the second derivative
of the energy functional.
\begin{equation}
\label{w_adiabatic}
w(x,x') \equiv \left.
  \frac{\delta^2 E[\rho]}{\delta\rho(x) \delta \rho(x')} \right|_{\rho_0} .
\end{equation}
Most of the energy functionals currently available are local in time,
which leads to $w(x,x') \propto \delta(t-t')$.

Multiplying the Dyson-type equation (\ref{Dyson_type_eq}) by the
perturbing external potential $v_1(x)$ leads to the linear density
response of Eq.~(\ref{rho_1}).
\begin{equation}
\label{rho_1_scf}
\rho_1(x)=\int d^4x' \Pi_s(x,x') v_{\rm scf}(x') ,
\end{equation}
where the self-consistent effective field, given by
\begin{equation}
\label{v_scf_t}
v_{\rm scf}(x) = v_1(x) + \int d^4y\ w(x,y) \rho_1(y) ,
\end{equation}
consists of the external perturbation $v_1$ and the induced residual field
$v_{\rm ks,1}(x)=\int d^4y w(x,y) \rho_1(y)$.
Thus, the {\it exact} representation of the linear density response $\rho_1(x)$
of a {\it real} interacting system can be written as the linear
density response of a {\it non-interacting} system to the self-consistent
effective perturbation $v_{\rm scf}(x)$.

The formal solution for the density response $\Pi$ is given by
solving the Dyson-type equation (\ref{Dyson_type_eq}),
$\Pi=(1-\Pi_s\cdot w)^{-1} \cdot \Pi_s$.
The non-interacting response $\Pi_s$ is explicitly given in
the followings, and
the residual kernel $w(x,x')$
is usually calculated using the adiabatic approximation
of Eq. (\ref{w_adiabatic}).
In this response function formalism,
the $\Pi(\omega)$ is usually solved in the frequency domain, to
calculate the strength function, transition density, etc.

\subsection{Linear density response with the Green's function method}
\label{sec: green_function}

The Fourier transform brings Eq. (\ref{rho_1_scf}) into
\begin{equation}
\label{rho_1_scf_FT}
\rho_1(\vec{r},\omega)=\int d^3r' \Pi_s(\vec{r},\vec{r'};\omega)
 v_{\rm scf}(\vec{r'},\omega) ,
\end{equation}
where the frequency-dependent effective field is given by
\begin{equation}
\label{v_scf}
v_{\rm scf}(\vec{r},\omega) = v_1(\vec{r},\omega) + \int d^3r'\
 w(\vec{r},\vec{r'}) \rho_1(\vec{r'},\omega) .
\end{equation}
The non-interacting response function $\Pi_s$ is expressed
in terms of the static Kohn-Sham orbitals
$\phi_k$ and their eigenenergies $\epsilon_k$:
\begin{equation}
\label{Pi_s_KS}
\begin{split}
\Pi_s(\vec{r},\vec{r'};\omega)
= \sum_{i\leq N} \phi_i^*(\vec{r}) \phi_i(\vec{r'})
   \sum_{m>N} \frac{\phi_m(\vec{r}) \phi_m^*(\vec{r'})}{\epsilon_i+\omega-\epsilon_m+i\eta} \\
+ \sum_{i\leq N} \phi_i(\vec{r}) \phi_i^*(\vec{r'})
   \sum_{m>N} \frac{\phi_m^*(\vec{r}) \phi_m(\vec{r'})}{\epsilon_i-\omega-\epsilon_m-i\eta} .
\end{split}
\end{equation}
The restriction for the summation with respect to the index $m$ can be
lifted, because the first and second terms in Eq. (\ref{Pi_s_KS}) give
the same magnitude but with an opposite sign for $m\leq N$.
Using the spectral representation of the one-particle retarded Green's function
for non-interacting particles,
\begin{equation}
G_s^{(+)}(\vec{r},\vec{r'};\omega)
 = \sum_k \frac{\phi_k(\vec{r})\phi_k^*(\vec{r'})}
                        {\omega-\epsilon_k+i\eta} ,
\end{equation}
one may replace summed orbitals with respect to $m$ 
in Eq. (\ref{Pi_s_KS}) by the
Green's function.
\begin{equation}
\label{Pi_s_Greenfn}
\Pi_s(\vec{r},\vec{r'};\omega) 
= \sum_{i\leq N} \left\{
  \phi_i^*(\vec{r}) \phi_i(\vec{r'})
  G^{(+)}(\vec{r},\vec{r'};\epsilon_i+\omega)
+ \phi_i(\vec{r}) \phi_i^*(\vec{r'})
  G^{(+)*}(\vec{r},\vec{r'};\epsilon_i-\omega)
  \right\} .
\end{equation}
This expression has practical advantages:
There is no need of truncation in the single-particle space,
as far as the Green's function is properly calculated.
Furthermore, the boundary condition imposed on the Green's function
provides the exact treatment of the continuum states \cite{SB75,ZS80}.
Normally, the energy of the occupied orbital is negative, $\epsilon_i<0$
($i=1,\cdots,N$).
Thus, the Green's function in the second term in  Eq.~(\ref{Pi_s_Greenfn})
always has a damped behavior in an asymptotic region ($r\rightarrow\infty$)
because of its negative argument $\epsilon_i-\omega<0$.
However, the first term may have an oscillatory behavior for
$\epsilon_i+\omega \geq 0$, which is provided by the outgoing boundary
condition in $G_s^{(+)}$.

An impulsive external potential associated with the function
$F(\vec{r})$,
\begin{equation}
\label{v_F}
v_1(x)=-s F(\vec{r})\delta(t) ,
\end{equation}
produces the density response $\rho_1(x)$ as Eq. (\ref{rho_1}).
Here, the parameter $s$ has the dimension of $ML^2T^{-1}[F(\vec{r})]^{-1}$.
The following quantity measures the collectivity of the response:
\begin{equation}
\label{R_F_t}
R_F(t) = \frac{-1}{s} \int d^3r F(\vec{r}) \rho_1(\vec{r},t) .
\end{equation}
The Fourier transform of $R_F(t)$ is given by
\begin{equation}
\label{R_F}
R_F(\omega) = \int R_F(t) e^{i\omega t} dt
 = \int d^3r \int d^3r'
F(\vec{r}) \Pi(\vec{r},\vec{r'};\omega) F(\vec{r'}) .
\end{equation}
Assuming $\omega\geq 0$ and using the relation
$(x+i\eta)^{-1}={\mathcal P} x^{-1} -i\pi\delta(x)$,
the strength function is obtained from the imaginary part of $R_F(\omega)$.
\begin{equation}
\label{S_F}
S_F(\omega)\equiv \sum_n |\bra{n} \hat{F} \ket{0}|^2
\delta(\omega-(E_n-E_0))
= \frac{-1}{\pi} {\rm Im} R_F(\omega) .
\end{equation}

\subsection{Real-time method}
\label{sec: real-time}

According to the response function formalism in Sec.~\ref{sec: green_function},
we need to construct the density-density response function $\Pi$ by
solving the Dyson-type equation (\ref{Dyson_type_eq}).
The required numerical task significantly increases as
the dimension of the response function
$\Pi(\vec{r},\vec{r'};\omega)$ increases,
and it has been practically prohibited for non-spherical systems.
In contrast,
the real-time method solving the TDKS equation in real time
provides a practical and efficient tool for calculation of
the strength function \cite{YB96,NY01,NY05,YKNI11}.
The method is based on the numerical integration of
the time evolution of the Kohn-Sham orbitals, described by
the TDKS equation (\ref{TDKS}).
The external perturbation $v(\vec{r},t)$ is taken to be small enough
to validate the linear approximation.
A good account of the continuum is given by
the use of complex absorbing potential  \cite{NY01,NY05}.
Recently, the canonical-basis real-time method has been developed
for open-shell nuclei with the BCS-like pairing.
This is based on the diagonal approximation for the time-dependent
pair potential \cite{Eba10}.
In this paper, we cannot present the details of these methods, but
readers should be referred to Refs.  \cite{NY01,NY05,Eba10}.

To calculate the density-density response function $\Pi$,
one must evaluate the residual kernel
$w(\vec{r},\vec{r'})=\delta v_{\rm ks}(\vec{r})/\delta\rho(\vec{r'})$
(Eq. (\ref{w_adiabatic}) in the adiabatic approximation),
which is a tedious task for realistic nuclear EDFs.
In real-time method, all we need to evaluate is the KS potential
$v_{\rm ks}[\rho(t)]$.
This is a practical advantage in the real-time method.
However, the real-time method often encounters a problem in numerical
stability.
This is especially serious in nuclear physics, because there is no
static external potential to hold the nuclear center of mass
at a fixed position.
Since the translational motion has no restoring force,
the moving system eventually hits the boundary of the space
and produces a spurious contribution to physical quantities.
Therefore, it is desirable to develop a practical methodology
in the frequency domain (representation), keeping
the advantageous features in the real-time method.
This is the finite amplitude method (FAM) \cite{NIY07},
which is presented in Sec.~\ref{sec: FAM}.

To illustrate the basic idea of the FAM,
in the followings,
we recapitulate the standard density matrix formulation and
its particle-hole representation.

\subsection{Matrix formulation in the particle-hole representation}
\label{sec: L-TDKS}

The most standard formulation of the density response is a matrix
formulation \cite{RS80,BR86}.
We start from the TDKS equation (\ref{TDKS_density}),
where $h_s(t)$ contains an external perturbation $v_1(t)$.
Provided that $v_1(t)$ is weak, we may linearize the TDKS equation
with respect to $v_1(t)$ and to the density response $\rho_1(t)$.
\begin{eqnarray}
\rho(t)&=&\rho_0+\rho_1(t) ,  \\
h_s(t)&=&h_0+v_{\rm scf}(t) ,
\end{eqnarray}
where $h_0=h_s[\rho_0]$ is the static KS Hamiltonian at the ground state
and $v_{\rm scf}(t)$ is a self-consistent effective field
induced by density fluctuations, Eq. (\ref{v_scf_t}):
\begin{equation}
v_{\rm scf}(t) 
 = v_1(t) + v_{\rm ks,1}(t),
\end{equation}
where $v_{\rm ks,1}(t)=w(t)\cdot \rho_1(t)
={\delta v_{\rm ks}}/{\delta \rho} \cdot \rho_1(t)$.
It should be noted that $v_{\rm ks,1}(t)$ has a linear dependence on
$\rho_1(t)$.
Using the stationary condition of the ground-state density,
$[h_0,\rho_0]=0$,
this leads to a time-dependent linear-response equation
with an external field,
\begin{equation}
\label{LTDKS}
i\frac{d}{dt}\rho_1(t) = [h_0,\rho_1(t)]
                       +[v_{\rm scf}(t),\rho_0] ,
\end{equation}
which can be written in the frequency domain as
\begin{equation}
\label{LRE}
\omega\ \rho_1(\omega) = [h_0,\rho_1(\omega)]
                       +[v_{\rm scf}(\omega),\rho_0] .
\end{equation}
Here, we decompose $\rho_1(t)$ and $v_{\rm scf}(t)$ into
those with fixed frequencies:
\begin{eqnarray}
\label{delta_rho_omega}
\rho_1(t)&=&\sum_\omega \big\{ \eta \rho_1(\omega) e^{-i\omega t}
                       +\eta^* \rho_1^\dagger(\omega) e^{i\omega t} \big\} ,\\
v_{\rm scf}(t)&=&\sum_\omega \left\{ \eta v_{\rm scf}(\omega) e^{-i\omega t} 
           + \eta^* v_{\rm scf}^\dagger (\omega) e^{i\omega t} \right\} ,
\end{eqnarray}
where we have introduced a small dimensionless parameter $\eta$.
$v_{\rm scf}(\omega)$ is a sum of $v_1(\omega)$ and
$v_{\rm ks,1}(\omega)=\delta v_{\rm ks}/\delta\rho \cdot \rho_1(\omega)$.
Note that the transition density $\rho_1(\omega)$,
the external field $v_1(\omega)$,
and the induced field $v_{\rm ks,1}(\omega)$,
are not necessarily Hermitian in the $\omega$-representation.

The time-dependent KS orbitals as solutions of Eq. (\ref{G-TDKS})
are written as $\ket{\psi_i(t)}=\ket{\phi_i}+\ket{\psi_{1,i}(t)}$,
where $\ket{\phi_i}$ are time-independent
eigenstates of the ground-state KS Hamiltonian
$ h_0 \ket{\phi_k} = \epsilon_k \ket{\phi_k}$.
A proper gauge choice $\xi_{ij}(t)$ should be adopted
to make the stationary eigenstates $\ket{\phi_i}$ consistent with
$\partial\ket{\psi_i}/\partial t=0$; e.g.,
$\xi_{ij}=\epsilon_i\delta_{ij}$.
Then, the time-dependent density response is
\begin{equation}
\rho_1(t)=\rho(t)-\rho_0 
=\sum_{i=1}^N \left\{
 \ket{\psi_{1,i}(t)}\bra{\phi_i} + \ket{\phi_i}\bra{(\psi_{1,i}(t)} \right\} .
\end{equation}
$\ket{\psi_{1,i}}$ are expanded in the Fourier components as
\begin{equation}
\label{psi_1}
\ket{\psi_{1,i}(t)}=
\sum_\omega \left\{ \eta \ket{X_i(\omega)} e^{-i\omega t} 
                          + \eta^* \ket{Y_i(\omega)} e^{i\omega t}  \right\} .
\end{equation}
and the density response at the frequency $\omega$ is given by
\begin{equation}
\label{rho_1_omega}
\rho_1(\omega) = \sum_i \big\{ \ket{X_i(\omega)}\bra{\phi_i}
                       + \ket{\phi_i}\bra{Y_i(\omega)} \big\} ,
\end{equation}
The orthonormalization of the TDKS orbitals leads to the fact that
only the particle-hole (ph) and hole-particle (hp)
matrix elements of $\rho_1(\omega)$ are non-zero.
Namely, $(\rho_1)_{ij}=(\rho_1)_{mn} = 0$ for $i,j \leq N$ and $m,n > N$.
Thus, without losing generality,
we can assume that the amplitudes, $\ket{X_i(\omega)}$ and $\ket{Y_i(\omega)}$,
can be expanded in the particle orbitals only;
\begin{equation}
\label{FB_amp}
\ket{X_i(\omega)}=\sum_{m>N} \ket{\phi_m} X_{mi}(\omega) ,
\quad\quad
\ket{Y_i(\omega)}=\sum_{m>N} \ket{\phi_m} Y_{mi}^*(\omega) .
\end{equation}
Using the $M\times N$ matrix
$\bar\varphi_{\alpha i}\equiv \inproduct{\alpha}{\phi_i}$ for the hole orbitals,
and the $M \times (M-N)$ matrix for the particle orbitals
$\varphi_{\alpha m} \equiv \inproduct{\alpha}{\phi_m}$,
the matrix $\rho_1(\omega)$ can be expressed by
\begin{equation}
\label{delta_rho_XY}
\rho_1(\omega)=\varphi X \bar\varphi^\dagger 
              +\bar\varphi Y^T \varphi^\dagger  .
\end{equation}
From this expression, it is apparent that the ph and
hp matrix elements of $\rho_1(\omega)$ represent
$X(\omega)$ and $Y(\omega)$, respectively.

If we take ph and hp matrix elements of Eq.~(\ref{LRE}),
we can derive the well-known linear response equation
in the matrix form  \cite{RS80};
\begin{equation}
\label{RPA}
\sum_{nj}
\left\{
\begin{pmatrix}
A   & B \\
B^* & A^*
\end{pmatrix}
- \omega
\begin{pmatrix}
1 & 0 \\
0 & -1
\end{pmatrix}
\right\}_{mi,nj}
\begin{pmatrix}
X_{nj}(\omega) \\
Y_{nj}(\omega)
\end{pmatrix}
= -
\begin{pmatrix}
f_{mi}(\omega) \\
g_{mi}(\omega)
\end{pmatrix} .
\end{equation}
Here, the matrices, $A$ and $B$, and the vectors, $f$ and $g$,
are defined by
\begin{eqnarray}
A_{mi,nj}&\equiv& (\epsilon_m-\epsilon_i)\delta_{mn}\delta_{ij}
+ w_{mj,in} , \quad\quad
B_{mi,nj}\equiv w_{mn,ij} ,\\
f_{mi}(\omega)&\equiv& \bra{\phi_m} v_1(\omega) \ket{\phi_i} ,\quad
g_{mi}(\omega)\equiv \bra{\phi_i} v_1(\omega) \ket{\phi_m} .
\end{eqnarray}
The residual kernel $w$ is often called Landau-Migdal interaction,
defined by
\begin{equation}
w_{mk,il}\equiv \bra{\phi_m}
\left.\frac{\partial v_{\rm ks}[\rho]}{\partial\rho_{lk}}\right|_{\rho=\rho_0}\ket{\phi_i} .\\
\end{equation}

In nuclear physics, this matrix formulation is also known as
the random-phase approximation (RPA).
The matrix ${\mathcal S}\equiv\begin{pmatrix} A & B\\ B^* & A^*\end{pmatrix}$
in Eq. (\ref{RPA}) is Hermitian and called ``stability matrix'' because its
eigenvalues characterize the stability of the ground state
determined by the solution of the KS(B) equations.
If ${\mathcal S}$ is positive definite, the ground state
is {\it stable}, thus corresponds to a (local) minimum.
In contrast, if ${\mathcal S}$ has negative eigenvalues,
the ground state with $\rho_0$ is not a minimum,
and there exists another true ground state.

In practical applications, the most tedious part is calculation of
the residual kernel, $w_{mj,ni}$ ($w_{mn,ij}$).
These elements are two-body-type matrix elements with four indices.
Their calculation is often the most demanding part in
numerical calculations.
In the next subsection,
we propose an alternative approach to
a solution of the linear-response equation (\ref{LRE}),
without the explicit evaluation of the residual kernel.

\subsection{Finite amplitude method}
\label{sec: FAM}

Let us remind ourselves that Eq.~(\ref{RPA})
was obtained by expanding $v_{\rm scf}(\omega)$ 
with respect to $X_{mi}(\omega)$ and $Y_{mi}(\omega)$.
The essential idea of the finite amplitude method (FAM) is to
perform this expansion implicitly in the numerical calculation.

Equation (\ref{RPA}) reads
\begin{equation}
\label{LRE_FAM}
\begin{split}
&(\epsilon_m-\epsilon_i-\omega) X_{mi}(\omega) 
+ (v_{\rm ks,1})_{mi}(\omega) = - (v_1)_{mi}(\omega) , \\
&(\epsilon_m-\epsilon_i+\omega) Y_{mi}(\omega) 
+ (v_{\rm ks,1})_{im}(\omega) = - (v_1)_{im}(\omega) ,
\end{split}
\end{equation}
where
$v_{\rm ks,1}(\omega)
=\delta v_{\rm ks}/\delta\rho \cdot \rho_1(\omega)$.
In the FAM, instead of performing the explicit expansion of
$v_{\rm ks,1}(\omega)$ with respect to $X$ and $Y$,
we resort to the numerical linearization.
Now, let us explain how to achieve it.

For given amplitudes $X$ and $Y$,
$v_{\rm ks,1}(\omega)$ can be numerically calculated using the finite
difference with respect to a small but finite real parameter $\eta$.
\begin{equation}
\label{FAM}
v_{\rm ks,1}(\omega) = \frac{1}{\eta}
   \left( v_{\rm ks}[\rho_\eta(\omega)] - v_{\rm ks}[\rho_0] \right) ,
\end{equation}
where 
$\rho_\eta(\omega) \equiv \rho_0 + \eta \rho_1(\omega)$.
The parameter $\eta$ should be chosen small so that the second and
higher-order terms in $v_{\rm ks}[\rho_\eta]$ are negligible.
In the first order in $\eta$, $\rho_\eta$
can be expressed by
$\rho_\eta(\omega) 
                 = \bar\psi'_\eta \bar\psi_\eta^\dagger$,
where
\begin{equation}
\bar\psi'_\eta = \bar\varphi + \eta \varphi X(\omega),\quad
\bar\psi_\eta^\dagger = \left(\bar\varphi + \eta \varphi Y^*(\omega) \right)^\dagger
\end{equation}
Namely, $v_{\rm ks}[\rho_\eta]$ in Eq. (\ref{FAM}) is evaluated
simply by replacing the ket states $\ket{\phi_i}$ with
$\ket{\phi_i}+\eta \sum_m \ket{\phi_m} X_{mi}(\omega)$,
and the bra states $\bra{\phi_i}$ with
$\bra{\phi_i}+\eta \sum_m \bra{\phi_m} Y_{mi}(\omega)$.
Regarding the KS potential $v_{\rm ks}$ as the functional of
$\bar\varphi$ and $\bar\varphi^\dagger$, Eq. (\ref{FAM}) is
rewritten as
\begin{equation}
\label{FAM_2}
v_{\rm ks,1}(\omega) = \frac{1}{\eta}
   \left( v_{\rm ks}[\bar\psi_\eta',\bar\psi_\eta^\dagger]
 - v_{\rm ks}[\bar\varphi,\bar\varphi^\dagger] \right) .
\end{equation}
The most advantageous feature of the present approach
is that it only requires calculation of the KS potential,
$v_{\rm ks}[\rho_\eta]$.
This should be included in the computer programs of the
static DFT calculations.
Only extra effort necessary is to estimate the KS potential
with different bra and ket single-particle states,
$\bar\psi^\dagger$ and $\bar\psi'$.
Therefore, a minor modification of the static DFT computer code
will provide a numerical solution of the linear density response.
This is the essence of the FAM.

Using these numerical differentiation,
both sides of Eq. (\ref{LRE_FAM})
can be easily obtained by calculating the ph
and hp matrix elements of the KS potential $v_{\rm ks}$.
Since 
these are inhomogeneous linear equations with respect to
$\ket{X_i(\omega)}$ and $\bra{Y_i(\omega)}$.
we can employ a well-established iterative method
for their solutions.
See Sec.~\ref{sec: FAM_application} for more details.

A typical numerical procedure is as follows:
(1) Fix the frequency $\omega$ and
assume initial vectors ($n=0$),
$X_{mi}^{(n)}(\omega)$ and $Y_{mi}^{(n)}(\omega)$.
(2) Update the vectors,
$X_{mi}^{(n+1)}(\omega)$ and $Y_{mi}^{(n+1)}(\omega)$,
    using the algorithm of an iterative method.
(3) Calculate the residual of Eq. (\ref{LRE_FAM}).
    If its magnitude is smaller than a given accuracy,
    stop the iteration.
    Otherwise, go back to the step (2).

To calculate the strength function with respect to the Hermitian
operator $F$, we should adopt $v_1(\omega)=F$.
After reaching the solution $(X_{mi},Y_{mi})$, the strength function
(\ref{S_F}) is obtained by
\begin{equation}
\label{S_F_2}
S_F(\omega)=\frac{-1}{\pi}{\rm Im} R_F(\omega) ,
\end{equation}
where
\begin{equation}
\label{R_F_2}
R_F(\omega)={\rm tr} [F \rho_1(\omega)]
=\sum_{mi} \left\{ F_{im} X_{mi}(\omega) + F_{mi} Y_{mi}(\omega) \right\} .
\end{equation}

\subsection{Quasiparticle formalism with pairing correlations and FAM}
\label{sec: FAM_QRPA}

In case that the pairing correlations play essential roles,
we extend the previous TDKS formalism to the TDKSB
formalism in Sec.~\ref{sec: TDKSB}.
This is straightforward.
Starting from the equation for the time-dependent generalized density matrix,
Eq. (\ref{TDKSB_density-2}), 
we follow the same procedure as that in Sec.~\ref{sec: L-TDKS}.
In this section, all the quantities must be defined in the body-fixed frame
rotating in the gauge space, which are expressed with the prime (')
in Sec.~\ref{sec: TDKSB}
\footnote{This corresponds to the ``moving-frame harmonic equation''
in Sec.~\ref{sec: MFHE}}.
We omit the primes here for simplicity.

The TDKSB Hamiltonian ${\mathcal H}_s$ contains the unperturbed
${\mathcal H}_s^0$,  a time-dependent external
 perturbation ${\mathcal V}_1(t)$, and induced field ${\mathcal H}_{s,1}(t)$.
\begin{equation}
{\mathcal H}_s[R](t)= {\mathcal H}_s^0 + {\mathcal H}_{s,1}(t) + {\mathcal V}_1(t) .
\end{equation}
This leads to the generalized density matrix
$R(t)= R_0 + R_1(t)$,
where $R_0$ is the ground-state density in the rotating frame.
Following the same arguments as in Sec.~\ref{sec: L-TDKS},
we may derive
the linearized TDKSB equation for the generalized density response
 $R_1(\omega)$,
\begin{equation}
\label{L-TDHFB-density-2}
\omega R_1(\omega) =
\left[ {\mathcal H}_s^0 -\mu {\mathcal N}, R_1(\omega) \right]
+ \left[ {\mathcal H}_{s,1}(\omega)+{\mathcal V}_1(\omega), R_0 \right] .
\end{equation}
Using the matrix notation of Eq. (\ref{Phi_matrix}),
the qp orbitals are expressed as\footnote{
Here, we assume a proper choice for the gauge parameter $\Xi_{\nu\nu'}$
to make a stationary solution $\bar\Phi$ time-independent.
}
\begin{equation}
\bar\Psi(t)=\bar\Phi + \bar\Psi_1(t) , \quad
\Psi(t)=\Phi + \Psi_1(t) , 
\end{equation}
with $R_0=\bar{\Phi}\bar{\Phi}^\dagger = 1-\Phi\Phi^\dagger$.
$\bar\Psi_1(t)$ ($\Psi_1(t)$) can be expanded only in terms of the
``unoccupied'' (``occupied'')
static orbitals $\Phi$ ($\bar\Phi$), thus written as
\begin{equation}
\label{Psi_1}
\begin{split}
&\bar\Psi_1(t)=\Phi \sum_\omega \left\{
X(\omega) e^{-i\omega t} + Y^*(\omega) e^{i\omega t} \right\} \\
&\Psi_1(t)=\bar\Phi \sum_\omega \left\{
X^*(\omega) e^{+i\omega t} + Y(\omega) e^{-i\omega t} \right\} .
\end{split}
\end{equation}
$X_{\nu\nu'}(\omega)$ and $Y_{\nu\nu'}(\omega)$ are $M\times M$ matrices,
which must be anti-symmetric because of the unitarity of
${\mathcal W}(t)=(\Psi(t),\bar\Psi(t))$.
The density $R(t)=\bar\Psi(t)\bar\Psi^\dagger(t)$ is
expanded up to the first order in $\bar\Psi_1$,  which gives
$
R_1(t)=
  \bar\Psi_1(t) \bar\Phi^\dagger
+ \bar\Phi \bar\Psi_1^\dagger(t)$.
Substituting Eq. (\ref{Psi_1}) into this, we have
\begin{equation}
R_1(\omega)= \Phi X \bar\Phi^\dagger + \bar\Phi Y^T \Phi^\dagger .
\end{equation}
From this expression, one can see that the ``unoccupied''-``occupied''
matrix elements of $R_1(\omega)$ are expressed by $X$, and
the ``occupied''-``unoccupied'' matrix elements are given by $Y$.
This is analogous to Eq. (\ref{delta_rho_XY}).
Using the unitarity of $W=(\Phi,\bar\Phi)$ and the following relations,
\begin{equation}
\left( {\mathcal H}_s^0-\mu{\mathcal N} \right) \Phi_\nu = E_\nu \Phi_\nu ,
\quad
\left( {\mathcal H}_s^0-\mu{\mathcal N} \right) \bar\Phi_\nu = -E_\nu \bar\Phi_\nu ,
\end{equation}
then, Eq. (\ref{L-TDHFB-density-2})
leads to the linear density response equations:
\begin{equation}
\label{LRE_QRPA}
\begin{split}
(E_\nu + E_{\nu'}-\omega ) X_{\nu\nu'} + ({\mathcal H}_{s,1})^{20}_{\nu\nu'}(\omega)
 = -({\mathcal V}_1)^{20}_{\nu\nu'}(\omega) , \\
(E_\nu + E_{\nu'}+\omega ) Y_{\nu\nu'} + ({\mathcal H}_{s,1})^{02}_{\nu\nu'}(\omega)
 = -({\mathcal V}_1)^{02}_{\nu\nu'}(\omega) ,
\end{split}
\end{equation}
where
\begin{eqnarray}
({\mathcal H}_{s,1})^{20}_{\nu\nu'}(\omega) &=&
  \left[ \Phi^\dagger {\mathcal H}_{s,1}(\omega) \bar\Phi \right]_{\nu\nu'} , \quad
 ({\mathcal V}_1)^{20}_{\nu\nu'}(\omega) =
  \left[ \Phi^\dagger {\mathcal V}_1(\omega) \bar\Phi \right]_{\nu\nu'} , \\
({\mathcal H}_{s,1})^{02}_{\nu\nu'}(\omega) &=&
  -\left[ \bar\Phi^\dagger {\mathcal H}_{s,1}(\omega) \Phi \right]_{\nu\nu'} , \quad
 ({\mathcal V}_1)^{02}_{\nu\nu'}(\omega) =
  -\left[ \bar\Phi^\dagger {\mathcal V}_1(\omega) \Phi \right]_{\nu\nu'} .
\end{eqnarray}
If we expand ${\mathcal H}_{s,1}^{20}$
and ${\mathcal H}_{s,1}^{02}$ in terms of $X$ and $Y$,
we reach the familiar expression of the matrix form, similar to 
Eq.~(\ref{RPA}).
The $A$ and $B$ matrices are given by the qp energy $E_\mu$ and
the residual kernels,
\begin{equation}
\label{QRPA_AB}
A_{\mu\nu,\mu'\nu'}=(E_\mu+E_\nu)\delta_{\mu\mu'} \delta_{\nu\nu'}
  +\frac{\partial ({\mathcal H}_s)_{\mu\nu}^{20}}
    {\partial R^{20}_{\mu'\nu'}} ,
\quad
B_{\mu\nu,\mu'\nu'}=
  \frac{\partial ({\mathcal H}_s)_{\mu\nu}^{20}}
    {\partial R^{02}_{\mu'\nu'}} ,
\end{equation}
where  the ``unoccupied''-``occupied'' and ``occupied''-``unoccupied''
components of the generalized density, $R^{20}$ and $R^{02}$, are defined by
\begin{equation}
R^{20}_{\mu\nu}= \left[\Phi^\dagger R \bar\Phi \right]_{\mu\nu} ,
\quad
R^{02}_{\mu\nu}= -\left[\bar\Phi^\dagger R \Phi \right]_{\mu\nu} .
\end{equation}

The finite amplitude method (FAM) for the qp density response
is presented in Ref.~ \cite{AN11}.
Here, we recapitulate the essential idea and the result.
Instead of calculating the residual kernels in Eq. (\ref{QRPA_AB}),
$({\mathcal H}_{s,1})^{20}$ and
$({\mathcal H}_{s,1})^{02}$ in Eq. (\ref{LRE_QRPA}) are
numerically obtained by the finite difference.
First, we define the $\eta$-density $R_\eta(\omega)$ as
\begin{equation}
R_\eta(\omega) = R_0 + \eta R_1(\omega)
               = \bar\Psi'_\eta(\omega) \bar\Psi_\eta^\dagger(\omega) ,
\end{equation}
where
\begin{equation}
\bar\Psi'_\eta(\omega) = \bar\Phi + \eta \Phi X(\omega) ,\quad
\bar\Psi_\eta^\dagger(\omega) = \left( \bar\Phi + \eta \Phi Y^*(\omega) \right)^\dagger.
\end{equation}
Then, the induced residual fields are given by the following FAM
formula:
\begin{equation}
{\mathcal H}_{s,1}^{20} =  \Phi^\dagger
\frac{{\mathcal H}_s[R_\eta] - {\mathcal H}_s[R_0]}{\eta}
\bar\Phi
, \quad
{\mathcal H}_{s,1}^{02} =  -\bar\Phi^\dagger
\frac{{\mathcal H}_s[R_\eta] - {\mathcal H}_s[R_0]}{\eta}
\Phi .
\end{equation}
Equivalently, the FAM formula can be written in terms of the qp orbitals as
\begin{eqnarray}
{\mathcal H}_{s,1}^{20} &=&  \Phi^\dagger
\frac{{\mathcal H}_s[\bar\Psi_\eta',\bar\Psi_\eta^\dagger]
 - {\mathcal H}_s[\bar\Phi,\bar\Phi^\dagger]}{\eta}
\bar\Phi \\
{\mathcal H}_{s,1}^{02} &=&  -\bar\Phi^\dagger
\frac{{\mathcal H}_s[\bar\Psi_\eta',\bar\Psi_\eta^\dagger]
 - {\mathcal H}_s[\bar\Phi,\bar\Phi^\dagger]}{\eta}
\Phi
\end{eqnarray}

A computer program for stationary solutions of the KSB equation
is able to construct the KSB Hamiltonian ${\mathcal H}_s[R]$ from
the qp orbitals $\bar\Phi$.
Thus, a small extension of the code to calculate ${\mathcal H}_s$
for different $\bar\Phi$ and $\bar\Phi^\dagger$
allows us to turn the static KSB code into the one for
the linear response calculation.
The FAM significantly reduces the programming task of developing
a new code \cite{AN11,Sto11}.
It turns out to save the enormous computational resources as well,
in linear-response calculations for deformed nuclei  \cite{Sto11}.

\section{Beyond perturbative regime: Large amplitude dynamics}
\label{sec: LACM}

Nuclei exhibit a variety of collective phenomena 
with large-amplitude and anharmonic nature in the low-energy region.
For instance, the nuclear fission is a typical example for such
a large-amplitude collective motion, that
a single nucleus is split into two or more smaller nuclei.
To describe these large-amplitude phenomena, we are aiming at
developing a practical theory to extract a few optimal canonical variables,
to describe the {\it slow collective} motion,
which are well decoupled from the other
{\it fast intrinsic} degrees of freedom.
Then, upon the obtained submanifold, the collective Hamiltonian is
constructed with microscopic
determination of the collective mass parameters and potentials,
to calculate observables in nuclear collective phenomena.

There have been extensive efforts in nuclear theory for such
purposes (See recent review papers \cite{DKW00,KMSTY01}).
In this article, we present a classical theory of the
large amplitude collective motion.
The contents in this section is mostly based on former
works \cite{DKW00,NWD99,MNM00,HNMM07}.

\subsection{Basic concepts of a decoupled collective submanifold}

As is shown in Appendix of Ref.  \cite{DKW00},
the TDKS(B) equations are
identical to the classical Hamilton's equations of motion
with the canonical variables $\{\xi^\alpha,\pi_\alpha\}_{\alpha=1,\cdots,N_c}$.
The number of independent variables $N_c$ are 
in the order of $M^2$ for the description of the TDKSB dynamics.
Since $M$ is in principle infinite without the truncation,
$N_c$ could be huge for description of the large-amplitude motion.
Thus, it is desirable to extract a few canonical variables which
are approximately decoupled from the other degrees of freedom.
These variables are supposed to describe decoupled collective
motion of the many-body system.
There are several equivalent ways to present the basic concepts and
formulation of the theory.

We assume the collective motion of interest is a slow
motion which allows us to truncate the classical Hamiltonian
under the expansion with respect to momenta.
Up to the second order in momenta $\pi$, the system is
described by the Hamiltonian
\begin{equation}
\label{H}
H(\xi,\pi)=\frac{1}{2} B^{\alpha\beta}(\xi) \pi_\alpha \pi_\beta + V(\xi) .
\end{equation}
The summation with respect to the repeated 
symbol for upper and lower indices is assumed, hereafter.
The reciprocal mass tensor $B^{\alpha\beta}$ is defined by
\begin{equation}
B^{\alpha\beta}=\left.\frac{\partial^2 H(\xi,\pi)}
{\partial \pi_\alpha \partial \pi_\beta}\right|_{\pi=0} .
\end{equation}
The mass tensor $B_{\alpha\beta}$
is defined by $B_{\alpha\beta} B^{\beta\gamma}= \delta _\alpha^\gamma$,
as the inverse matrix of $B^{\alpha\beta}$.
We are trying to find a collective submanifold present in
the classical Hamilton system described by $H$ in the form
of Eq. (\ref{H}).

\subsubsection{Point transformation}

In general, The main aim of the theory is to find the canonical
transformation
\begin{equation}
\{ \xi^\alpha,\pi_\alpha \}
 \rightarrow  \{ q^\mu, p_\mu \} ,
\end{equation}
where
the $\{ q^\mu \}$ are assumed to be divided into two subsets,
$q^i$, $i=1,\cdots,K$ and the rest $q^a$, $a=K+1,\cdots,N_c$,
which are decoupled with each other.
Namely, if the system is located at $q^a=0$ and $\dot{q}^a=0$
at time $t=0$, then the time evolution should keep $q^a(t)=0$.
Of course, in reality, the decoupling is not exact.
We want to find an approximately decoupled manifold.

First, we limit ourselves to the point transformations.
\begin{equation}
\label{point_transf}
q^\mu=f^\mu(\xi), \quad \xi^\alpha=g^\alpha(q) .
\end{equation}
In the point transformation, the transformation for conjugate momenta
are given by
derivatives of the functions $f^\mu$ and $g^\alpha$.
\begin{equation}
\label{momentum_transf}
p_\mu=g^\alpha_{,\mu} \pi_\alpha, \quad \pi_\alpha=f^\mu_{,\alpha} p_\mu ,
\end{equation}
where the comma indicates a partial derivative,
$g^\alpha_{,\mu} = \partial g^\alpha/\partial q^\mu$.
The canonicity is guaranteed by the conservation of the Poisson
brackets, which is easily proven by using the chain-rule relations:
\begin{equation}
g^\alpha_{,\mu} f^\mu_{,\beta} = \delta^\alpha_\beta,
\quad
f^\mu_{,\alpha} g^\alpha_{,\nu} = \delta^\mu_\nu .
\end{equation}
Substituting the point transformation of Eq. (\ref{point_transf}) into
Eq. (\ref{H}), the Hamiltonian in the new variables becomes
\begin{equation}
\label{Hbar}
\bar{H}(q,p) = 
\frac{1}{2} \bar{B}^{\mu\nu}(q) p_\mu p_\nu + \bar{V}(q) .
\end{equation}
The reciprocal mass parameter transforms like a tensor of the second rank.
\begin{equation}
\bar{B}^{\mu\nu}(q) = f^\mu_{,\alpha} B^{\alpha\beta} f^\nu_{,\beta} .
\end{equation}

\subsubsection{Decoupling condition under a point transformation}

The decoupling condition is that, if the system is located on the
collective submanifold ($q^a=p^a=0$),
it stays within the submanifold,
namely, $\dot{q}^a=\dot{p}^a=0$.
From the Hamilton's equations of motion derived from Eq. (\ref{Hbar}),
\begin{equation}
\begin{split}
&\dot{q}^a = \frac{\partial \bar{H}}{\partial p^a}
          = \bar{B}^{ai} p_i + \bar{B}^{ab}p_b ,\\ 
&\dot{p}^a = -\frac{\partial \bar{H}}{\partial q^a}
          = -\bar{V}_{,a} - \frac{1}{2}\bar{B}^{\mu\nu}_{,a} p_\mu p_\nu ,
\end{split}
\end{equation}
we have the following conditions for the decoupling:
\begin{eqnarray}
\label{B_dc}
\bar{B}^{ai}=0 ,\\
\label{V_dc}
\bar{V}_{,a}=0 ,\\
\label{g_dc}
\bar{B}^{ij}_{,a}=0 .
\end{eqnarray}
The first condition, Eq. (\ref{B_dc}), tells us that the reciprocal mass tensor
must be block diagonal and has no coupling between the collective
space ($q^i$ with $i=1,\cdots,K$) and the intrinsic space
($q^a$ with $a=K+1,\cdots,N_c$).
The remaining two conditions comes from the absence of
the force perpendicular to the collective surface.
The conditions for the mass tensor, Eqs. (\ref{B_dc}) and (\ref{g_dc}),
imply that the decoupled submanifold is geodesic with
the metric given by the mass tensor $\bar{B}_{ij}$.
Namely, the following quantity
\begin{equation}
A_K = \int \sqrt{B} dq^1 \wedge \cdots \wedge dq^K ,
\quad B\equiv \det(\bar{B}_{ij}),
\end{equation}
with a fixed boundary $\partial S_K$ has a minimum value,
$\delta A_K =0$.
See Ref.  \cite{DKW00} for the proof.

Utilizing the chain rule,
the force condition, Eq.~(\ref{V_dc}),
can be rewritten as
\begin{equation}
\label{V_dc_Kdim}
V_{,\alpha}=\bar{V}_{,\mu}f^\mu_{,\alpha} = \bar{V}_{,i} f^i_{,\alpha} .
\end{equation}
This is the condition obtained in the zero-th order in momenta.
In the one-dimensional case ($K=1$), it is
\begin{equation}
\label{V_dc_1dim}
V_{,\alpha}-\lambda_1 f^1_{,\alpha}=0 ,
\end{equation}
where
$\lambda_1\equiv {\partial\bar{V}}/{\partial q^1}$.
This is nothing but the minimization of the potential $V(\xi)$ with
a constraint on the collective coordinate $q^1=f^1$.
\begin{equation}
\delta \left\{ V(\xi) - \lambda_1 f^1(\xi) \right\} = 0 .
\end{equation}

\subsection{Local harmonic equations (LHE)}
\label{sec: LHE}

If all of the decoupling conditions, Eqs. (\ref{B_dc}), (\ref{V_dc}), and
(\ref{g_dc}), are satisfied, it provides an exactly decoupled
collective submanifold.
However, in realistic situation, the exact decoupling is not realized
except for trivial collective motions, such as the translational motion.
We are more interested in situations with approximate decoupling.
Among the three decoupling conditions, Eq. (\ref{g_dc}) comes from
the coefficients in the second order in momenta.
Here, we build the theory by ignoring this second-order condition,
based on the mass condition (\ref{B_dc}) and
the force condition (\ref{V_dc}).

Let us start from the chain-rule about the derivative of the potential,
\begin{equation}
\label{V_derivative}
V_{,\alpha} = \bar{V}_{,\nu} f^\nu_{,\alpha} ,
\quad
\bar{V}_{,\mu} = V_{,\alpha} g^\alpha_{,\mu} ,
\end{equation}
This indicates that the first derivative has a property
of the covariant vector.
However, the second derivatives are known to be not a tensor with respect
to the general point transformation.
As is well-known in the general relativity,
we should introduce
the covariant derivative, to keep the tensorial property.
The covariant derivative is defined by
\begin{equation}
V_{;\alpha\beta} \equiv \lim_{d\xi \rightarrow 0}
   \frac{dV_{,\alpha}-\delta V_{,\alpha}} {d\xi^\beta}
= V_{,\alpha\beta}
     - \Gamma^\gamma_{\alpha\beta} V_{,\gamma} ,
\end{equation}
using the parallel transport of
the vector $V_{,\alpha}(\xi)$ for $\xi \rightarrow \xi + d\xi$,
\begin{equation}
    \delta V_{,\alpha}= \Gamma^\beta_{\alpha\gamma} V_{,\beta} d\xi^\gamma .
\end{equation}
In order to make the covariant derivatives $V_{;\alpha\beta}$ a tensor
of the second rank, the affine connection $\Gamma^\alpha_{\beta\gamma}$
must follow the transformation:
\begin{eqnarray}
\bar{\Gamma}^\mu_{\nu\rho} &=&
 f^\mu_{,\alpha} g^\beta_{,\nu} g^\gamma_{,\rho} \Gamma^\alpha_{\beta\gamma}
      +f^\mu_{,\alpha} g^\alpha_{,\nu\rho} , \\
\Gamma^\alpha_{\beta\gamma} &=&
 g^\alpha_{,\mu} f^\nu_{,\beta} f^\rho_{,\gamma} \bar\Gamma^\mu_{\nu\rho}
      +g^\alpha_{,\mu} f^\mu_{,\beta\gamma} .
\end{eqnarray}
Now, we assume that the coordinate system $\{ q^\mu \}$ is geodesic,
namely, $\bar{\Gamma}^\mu_{\nu\rho}=0$.
This leads to the affine connection
\begin{equation}
\label{symplectic_connection}
\Gamma^\alpha_{\beta\gamma} \equiv g^\alpha_{,\mu} f^\mu_{,\beta\gamma} ,
\end{equation}
and the covariant derivatives
\begin{equation}
\label{covariant_der}
V_{;\alpha\beta} \equiv V_{,\alpha\beta}
     - \Gamma^\gamma_{\alpha\beta} V_{,\gamma}
       = V_{,\alpha\beta}
     - f^\mu_{,\alpha\beta} \bar{V}_{,\mu} ,
\quad \quad
\bar{V}_{;\mu\nu} = \bar{V}_{,\mu\nu} ,
\end{equation}
which can be even simplified
because of Eq. (\ref{V_dc}), as
\begin{equation}
\label{covariant_der_2}
V_{;\alpha\beta} 
       = V_{,\alpha\beta}
     - f^i_{,\alpha\beta} \bar{V}_{,i} .
\end{equation}
Since these covariant derivatives are tensor,
they must transform as
\begin{equation}
V_{;\alpha\beta} = \bar{V}_{;\mu\nu} f^\mu_{,\alpha} f^\nu_{,\beta} .
\end{equation}
Multiplying the reciprocal mass tensor, we have
\begin{equation}
{\mathcal M}^\alpha_\beta \equiv
B^{\alpha\gamma}V_{;\gamma\beta}
 = \bar{B}^{\mu\rho}\bar{V}_{,\rho\nu} g^\alpha_{,\mu} f^\nu_{,\beta}
 = \bar{\mathcal M}^\mu_\nu g^\alpha_{,\mu} f^\nu_{,\beta} ,
\end{equation}
from which we easily obtain the following equations.
\begin{equation}
\label{M}
{\mathcal M}^\alpha_\beta f^\mu_{,\alpha}
 = \bar{\mathcal M}^\mu_\nu f^\nu_{,\beta} ,
\quad\quad
{\mathcal M}^\alpha_\beta g^\beta_{,\mu}
 = \bar{\mathcal M}^\nu_\mu g^\alpha_{,\nu} .
\end{equation}

Now, let us use the decoupling conditions, Eqs. (\ref{B_dc}) and (\ref{V_dc}).
Taking $\mu=i$ (collective coordinate) in Eq. (\ref{M}), 
the decoupling conditions tell us that the matrix
${\mathcal M}^\mu_\nu=\bar{B}^{\mu\rho}\bar{V}_{,\rho\nu}$
is also block diagonal,
${\mathcal M}^i_a=0$ and ${\mathcal M}^a_i=0$.
Therefore, we reach the following equations, which we call
``local harmonic equations'' (LHE).
\begin{equation}
\label{LHE}
{\mathcal M}^\alpha_\beta f^i_{,\alpha}
 = \bar{\mathcal M}^i_j f^j_{,\beta} ,
\quad\quad
{\mathcal M}^\alpha_\beta g^\beta_{,i}
 = \bar{\mathcal M}^j_i g^\alpha_{,j} .
\end{equation}
In the case of $K=1$, it is written as
\begin{equation}
\label{LHE_1}
{\mathcal M}^\alpha_\beta f^1_{,\alpha}
 =  \omega^2 f^1_{,\beta} ,
\quad\quad
{\mathcal M}^\alpha_\beta g^\beta_{,i}
 = \omega^2 g^\alpha_{,1} .
\end{equation}
where the frequency is given by $\omega^2=\bar{\mathcal M}^1_1$.
The solution of the LHE provides a tangent vector of the collective
submanifold, $f^i_{,\alpha}$ and $g^\alpha_{,i}$.

The LHE generalizes the secular equation of the harmonic approximation
around the potential minimum to that at non-equilibrium points.
At the equilibrium ($V_\alpha=\bar{V}_{,\mu}=0$),
it automatically reduces to
the normal harmonic approximation, because the covariant derivatives
become identical to the second derivatives at the equilibrium,
$V_{;\alpha\beta}=V_{,\alpha\beta}$.

\subsubsection{Practical solution of LHE}
\label{sec: practical}

To solve the LHE (\ref{LHE}),
we need to calculate the affine connection of Eq. (\ref{symplectic_connection}),
which contains the curvature $f^i_{,\alpha\beta}$, in the covariant derivative
$V_{;\alpha\beta}$.
Since the solution of the LHE provides only $f^i_{,\alpha}$ and
$g^\alpha_{,i}$,
this cannot be given by the LHE itself.
However, the curvature $f^i_{,\alpha\beta}$ can be eliminated
in the following procedure \cite{MNM00}.
Here, let us discuss the case of $K=1$, for simplicity.
In this case, we have a single collective coordinate $q^1$.
We take the derivative of
$\bar{B}^{11}=B^{\alpha\beta} f^1_{,\alpha} f^1_{,\beta}$
with respect to $q^1$.
\begin{equation}
\bar{B}_{,1}^{11}=
B^{\alpha\gamma}_{,\beta} f^1_{,\alpha} f^1_{,\gamma} g^\beta_{,1}
     + 2 B^{\alpha\gamma} f^1_{,\alpha} f^1_{,\gamma\beta} g^\beta_{,1} .
\end{equation}
Using this equation, we may eliminate the curvature terms in 
the LHE (\ref{LHE_1}).
The LHE can be rewritten in the same form as Eq. (\ref{LHE_1}), but
${\mathcal M}^\alpha_\beta$ and $\omega^2$ can be replaced by
\begin{equation}
\begin{split}
{\mathcal M}^\alpha_\beta &\equiv B^{\alpha\gamma} V_{,\gamma\beta}
 + \frac{1}{2} B^{\alpha\gamma}_{,\beta} V_{,\gamma} \\
\omega^2 &\equiv \bar{B}^{11} \bar{V}_{,11} + \frac{1}{2} \bar{B}^{11}_{,1} \bar{V}_{,1} .
\end{split}
\end{equation}
In this way, we can eliminate the curvature terms.
The price to pay is the calculation of $B^{\alpha\gamma}_{,\beta}$.
The eigenfrequency $\omega^2$ is obtained by solving the eigenvalue
equation (\ref{LHE_1}).
Thus, We do not need to calculate $\bar{B}^{11}_{,1}$.

\subsubsection{Riemannian connection}

In this article, we adopt the affine connection of
Eq.~(\ref{symplectic_connection}), which assumes that the decoupled
coordinates $\{ q^\mu \}$ is geodesic ($\bar\Gamma^\mu_{\nu\rho}=0$).
Instead of Eq.~(\ref{symplectic_connection}),
the Riemannian connection may be adopted,
in a similar manner to the general relativity.
The Riemannian connection
is given in terms of the metric tensor $B_{\alpha\beta}$ as
\begin{equation}
\label{Riemannian_connection}
\Gamma^\alpha_{\beta\gamma} = \frac{1}{2} B^{\alpha\delta} \left(
B_{\beta\delta,\gamma}+B_{\gamma\delta,\beta} - B_{\beta\gamma,\delta}
\right) .
\end{equation}
In Ref.  \cite{DKW00}, this was discussed in details.
In fact, the decoupling conditions,
Eqs. (\ref{B_dc}), (\ref{V_dc}), and (\ref{g_dc}),
may lead to the LHE identical to Eq. (\ref{LHE})
with the connection $\Gamma^\alpha_{\beta\gamma}$ replaced by
Eq. (\ref{Riemannian_connection}).
However, the Riemannian formulation of the LHE has a problem
for the case that the Nambu-Goldstone modes exist  \cite{NWD99},
which will be discussed in Sec.~\ref{sec: NG_mode}
Thus, in the followings, we focus our discussion on the LHE
with the affine connection of Eq. (\ref{symplectic_connection}).

\subsection{Treatment of constants of motion: Nambu-Goldstone modes}
\label{sec: NG_mode}

Since nuclei are self-bound system without an external potential,
the nuclear DFT provides a ground-state density distribution
which spontaneously violates the symmetry, such as translational and rotational
symmetries.
The spontaneous breaking of symmetry produces the Nambu-Goldstone (NG)
modes which correspond to trivial (spurious) collective degrees of
freedom.
Therefore, we are mostly interested in the extraction of the
collective degrees of freedom which are separated from (perpendicular to)
these NG modes.
In this section, we show that the LHE presented in Sec.~\ref{sec: LHE}
properly separate the NG degrees of freedom from other degrees of
freedom.
However, to achieve this, we need to lift the restriction to the
point transformation and extend the point transformation to allow
the second-order terms in momenta \cite{NWD99}.

\subsubsection{Extended adiabatic transformation}

The restriction to the point transformation is lifted by
expansion with respect to momenta $\pi$.
Equations (\ref{point_transf}) are generalized by
\begin{eqnarray}
\label{EAT_0}
q^\mu &=& f^\mu(\xi)
  + \frac{1}{2} f^{(1)\mu\alpha\beta}(\xi) \pi_\alpha \pi_\beta
  + {\mathcal O}(\pi^4) , \\
\label{EAT_0b}
\xi^\alpha &=& g^\alpha(q)
  + \frac{1}{2} g^{(1)\alpha\mu\nu}(q) p_\mu p_\nu
  + {\mathcal O}(p^4) .
\end{eqnarray}
The transformation of the momenta is given by Eq. (\ref{momentum_transf}),
since the terms cubic in momenta do not play a role in the modification
of the theory.
Using Eq.~(\ref{momentum_transf}),
the independence of the variables, $\partial\xi^\alpha/\partial\pi_\beta=0$,
requires the relation
\begin{equation}
\label{EAT_1}
g^{(1)\alpha\mu\nu} g^\beta_{,\mu} g^\gamma_{,\nu} =
 -f^{(1)\lambda\beta\gamma} g^\alpha_{,\lambda} .
\end{equation}
From the canonicity condition $\{ q^\mu, q^\nu \}_{\rm PB} = 0$,
we also find
\begin{equation}
\label{EAT_2}
f^\mu_{,\alpha} f^{(1)\nu\alpha\beta}
 = f^\nu_{,\alpha} f^{(1)\mu\alpha\beta} .
\end{equation}

The Hamiltonian (\ref{H}) is transformed to,
up to second order in $p$,
\begin{equation}
\bar{H}(q,p) = \bar{V}(q) + \frac{1}{2} \bar{B}^{\mu\nu} p_\mu p_\nu ,
\quad\quad
\bar{B}^{\mu\nu} = f^\mu_{,\alpha} B^{\alpha\beta} f^\nu_{,\beta}
                     + V_{,\gamma} g^{(1)\gamma\mu\nu} .
\end{equation}
The major difference between the use of the extended adiabatic
transformation and a point transformation is the modification
of mass parameter,
\begin{eqnarray}
\label{B_tilde_1}
\widetilde{B}^{\alpha\beta} &\equiv&
   g^\alpha_{,\mu} \bar{B}^{\mu\nu} g^\beta_{,\nu}
 = B^{\alpha\beta} - \bar{V}_{,\mu} f^{(1)\mu\alpha\beta} \\
\label{B_tilde}
 &=& B^{\alpha\beta} - \bar{V}_{,i} f^{(1)i\alpha\beta} .
\end{eqnarray}
Here we have used Eqs. (\ref{EAT_1}) and (\ref{V_dc}).
The LHE has the same form as the Eq. (\ref{LHE}),
after replacing 
$B^{\alpha\beta}$ by $\widetilde{B}^{\alpha\beta}$.

\subsubsection{Constants of motion; cyclic variables}

Suppose a classical variable ${\mathcal P}(\xi,\pi)$,
which correspond to one-body Hermitian operators $P$ in the quantum mechanics,
is a constant of motion.
In the followings,
the conserved quantities are classified into two categories.
Adopting the classical canonical variables in Ref.~ \cite{NWD99},
if $P$ has real matrix elements in the qp basis,
${\mathcal P}$ can be expanded as
\begin{equation}
{\mathcal P}(\xi,\pi) = {\mathcal P}^{(0)}(\xi)
 + \frac{1}{2} {\mathcal P}^{(2)\alpha\beta} \pi_\alpha \pi_\beta
 + {\mathcal O}(\pi^4)  .
\end{equation}
On the other hand, if $P$ has imaginary matrix elements,
\begin{equation}
{\mathcal P}(\xi,\pi) = {\mathcal P}^{(1)\alpha} \pi_\alpha
 + {\mathcal O}(\pi^3)  .
\end{equation}
The conservation of ${\mathcal P}$ indicates that
the Poisson bracket between ${\mathcal P}$ and $H$ should vanish.
From this, terms of the zeroth and first order in $\pi$ give
the following identities.
\begin{eqnarray}
\label{conserved_0}
{\mathcal P}^{(1)\alpha} V_{,\alpha} = 0 ,\\
\label{conserved_1}
{\mathcal P}^{(0)}_{,\alpha} B^{\alpha\beta}
 - {\mathcal P}^{(2)\alpha\beta} V_{,\alpha} = 0 .
\end{eqnarray}
The equations (\ref{conserved_0}) and (\ref{conserved_1})
hold at arbitrary points in the configuration space.

We assume that the variables describing these constants of motion
correspond to the canonical variables $(q^I,p_I)$.
The collective variables $(q^i,p_i)$ of interest are supposed to
be orthogonal to both these variables and the intrinsic variables $(q^a,p_a)$.
Thus, we divide the set $\{q^\mu,p_\mu\}$ ($\mu=1,\cdots,N_c$)
into three subsets,
the collective coordinates $\{q^i,p_i\}$, $i=1,\cdots,K$,
the cyclic coordinates $\{q^I,p_I\}$, $I=K+1,\cdots,K+M$, and
the non-collective coordinates $\{q^a,p_a\}$, $a=K+M+1,\cdots,N_c$.

In nuclear physics applications, we are often interested in
the large amplitude collective motion at a given value of $q^I$,
such as the given total angular momentum,
$q^I=\langle J_x \rangle=J$, and
the given number of particles in the superfluid systems,
 $q^I=\langle N \rangle=N$.
In this case, 
Eq. (\ref{V_dc_Kdim}) should be modified with additional constraints
with respect to $q^I$ as
\begin{equation}
\label{V_dc_NG}
V_{,\alpha}-\bar{V}_{,i} f^i_{,\alpha} - \bar{V}_{,I} f^I_{,\alpha} = 0 .
\end{equation}
Here, $f^I_{,\alpha}$ are given by
the 2qp matrix elements of the symmetry operator.
The non-trivial collective coordinates of interest, $f^i_{,\alpha}$,
are determined by the solution of the LHE (\ref{LHE}) 
with the reciprocal mass tensor of Eq. (\ref{B_tilde}).
We should solve Eqs. (\ref{V_dc_NG}) and (\ref{LHE}) self-consistently.

\subsubsection{Separation of cyclic variables as zero modes}

Now, let us prove that $f^I_{,\alpha}$ ($g^\alpha_{,I}$)
provides the zero-frequency solution ($\omega=0$) for the LHE.
We start from the case that the coordinates $q^I$ are conserved,
with $f^I(\xi)={\mathcal P}^{(0)}(\xi)$ and
 $f^{(1)I\alpha\beta(\xi)}={\mathcal P}^{(2)\alpha\beta}(\xi)$.
This corresponds to the case of most practical interests in nuclear physics,
such as the angular momentum and particle number.
\begin{equation}
\begin{split}
{\mathcal M}^\alpha_\beta f^I_{,\alpha} &=
\widetilde{B}^{\alpha\gamma} V_{;\gamma\beta} f^I_{,\alpha} 
= (B^{\alpha\gamma} f^I_{,\alpha}
 - \bar{V}_{,\mu} f^{(1)\mu\alpha\gamma} f^I_{,\alpha} ) V_{;\gamma\beta} \\
&= (B^{\alpha\gamma} f^I_{,\alpha}
 - \bar{V}_{,\mu} f^{(1)I\alpha\gamma} f^\mu_{,\alpha} ) V_{;\gamma\beta}
= (B^{\alpha\gamma} f^I_{,\alpha}
 - V_{,\alpha} f^{(1)I\alpha\gamma} ) V_{;\gamma\beta}
=0 .
\end{split}
\end{equation}
Here, Eq. (\ref{EAT_2}) was used in the third equation, and
Eq. (\ref{conserved_1}) was in the last equation.
Thus, $f^I_{,\alpha}$ automatically becomes
a solution of the LHE with $\omega=0$.

Next, we discuss the case that the momenta $p_I$ are conserved,
with $g^I_{,\alpha}(\xi)={\mathcal P}^{(1)\alpha}(\xi)$.
Differentiating the chain relation
$g^\alpha_{,\mu} f^\mu_{,\beta} = \delta^\alpha_\beta$
with respect to $q^\nu$, we obtain
\begin{equation}
g^\alpha_{,\mu\nu} f^\mu_{,\beta}
  = - g^\alpha_{,\mu} g^\gamma_{,\nu} f^\mu_{,\beta\gamma} .
\end{equation}
Differentiating Eq. (\ref{conserved_0}), we have
\begin{equation}
V_{,\alpha\beta}g^\alpha_{,I} + V_{,\alpha}g^\alpha_{,I\mu} f^\mu_{,\beta}=0.
\end{equation}
Utilizing these equation, we may prove
\begin{equation}
\begin{split}
{\mathcal M}^\alpha_\beta g^\beta_{,I} &=
\widetilde{B}^{\alpha\gamma} V_{;\gamma\beta} g^\beta_{,I} 
=\widetilde{B}^{\alpha\gamma} (V_{,\gamma\beta} -f^\mu_{,\gamma\beta} \bar{V}_{,\mu} )
 g^\beta_{,I}  \\
&=\widetilde{B}^{\alpha\gamma} (V_{,\gamma\beta} g^\beta_{,I}
-f^\mu_{,\gamma\beta}
 V_{,\delta} g^\delta_{,\mu} g^\beta_{,I} )
=\widetilde{B}^{\alpha\gamma} (V_{,\gamma\beta} g^\beta_{,I}
 +f^\mu_{,\gamma}  g^\delta_{,I\mu} V_{,\delta} ) 
=0 .
\end{split}
\end{equation}
Therefore, the $g^\alpha_I$ are zero-frequency solutions of
the LHE.

The separation of the NG modes are guaranteed in the LHE
with the covariant derivatives $V_{;\alpha\beta}$
of Eq. (\ref{covariant_der_2}) and the reciprocal mass tensor
$\widetilde{B}^{\alpha\beta}$ of Eq.~(\ref{B_tilde}).

\subsection{Gauge invariance}

The basic formulation to determine the collective submanifold is
given by Eqs. (\ref{V_dc_Kdim}) and (\ref{LHE}).
In the case of the one-dimensional collective coordinate ($K=1$),
these equations provide a unique solution, except for the scale
of the collective coordinate $q^1$.
However, for the multi-dimensional collective manifold ($K>0$),
the solution of Eqs. (\ref{V_dc_Kdim}) and (\ref{LHE})
are not unique.
In fact, there is a gauge invariance similar to what we observed
in Eqs. (\ref{G-TDKS}) and (\ref{G-TDKSB}).
For a pair of collective variables $(q^k,p_k)$ and $(q^l,p_l)$,
$k\neq l$, we may adopt a point transformation
\begin{equation}
\label{gauge_transf}
q^k \rightarrow q^k + c q^l ,
\quad\quad
p_l \rightarrow p_l - c p_k ,
\end{equation}
with an arbitrary gauge parameter $c$, keeping the other variables
unchanged.
Let us show the transformation of Eq.~(\ref{gauge_transf}) keeps
the formulation of Eqs. (\ref{V_dc_Kdim}) and (\ref{LHE}) invariant.
Namely, $(q^k+cq^l,p_k)$ and $(q^l,p_l-cp_k)$ instead of
$(q^k,p_k)$ and $(q^l,p_l)$ also provides a self-consistent solution
for Eqs. (\ref{V_dc_Kdim}) and (\ref{LHE}).

Since the transformation (\ref{gauge_transf}) gives
$g^\alpha_l \rightarrow g^\alpha_{,l} -c g^\alpha_{,k}$,
the derivative of the potential, $\bar{V}_{,l}=V_{,\alpha}g^\alpha_{,l}$,
is transformed as
$ \bar{V}_{,l} \quad
\rightarrow \quad
\bar{V}_{,l} -c \bar{V}_{,k}
$.
From this, we can immediately see
the invariance of Eq. (\ref{V_dc_Kdim}), using the
transformation $f^k_{,\alpha}\rightarrow f^k_{,\alpha} + c f^l_{,\alpha}$.
The matrix
${\mathcal M}^\alpha_\beta$ in the left-hand side of Eq. (\ref{LHE})
is also invariant under Eq. (\ref{gauge_transf}).
In fact, both $\widetilde{B}^{\alpha\beta}$ and $V_{;\alpha\beta}$
are separately invariant.
In contrast, 
the matrix $\bar{\mathcal M}^{i}_{j}$ in the right-hand side of
Eq. (\ref{LHE}) transforms as
\begin{equation}
\begin{split}
&\bar{\mathcal M}^k_i \rightarrow \bar{\mathcal M}^k_i + c\bar{\mathcal M}^l_i ,
\quad\quad
\bar{\mathcal M}^j_l \rightarrow \bar{\mathcal M}^j_l - c \bar{\mathcal M}^j_k ,
\quad\quad
\bar{\mathcal M}^j_i \rightarrow \bar{\mathcal M}^j_i ,\\
&\bar{\mathcal M}^k_l \rightarrow \bar{\mathcal M}^k_l + c\bar{\mathcal M}^l_l
                        - c \bar{\mathcal M}^k_k - c^2 \bar{\mathcal M}^l_k
\end{split}
\end{equation}
for $i\neq l$ and $j\neq k$.
This can be easily obtained from the relation
 $\bar{\mathcal M}^i_j= f^i_{,\alpha} {\mathcal M}^\alpha_\beta g^\beta_{,j}$.
These relations prove that
$\{ f^i_{,\alpha} \}_{i=1,\cdots,K}$ with $f^k_{,\alpha}$
 replaced by $f^k_{,\alpha}+cf^l_{,\alpha}$
also provides a solution of Eq. (\ref{LHE}).
In the same manner, we can prove that 
$\{ g^\alpha_{,i} \}_{i=1,\cdots,K}$ with $g^\alpha_{,l}$
replaced by $g^\alpha_{,l}-cg^\alpha_{,k}$
is a solution as well.
This gauge invariance is present for any pair of collective variables
$(k,l)$, thus for an arbitrary linear point transformation.

In the case that the cyclic variables $(q^I,p_I)$ exist,
the gauge invariance is present even for $K=1$.
Suppose $q^1$ is a collective coordinate, which is a self-consistent solution
of Eqs. (\ref{V_dc_NG}) and (\ref{LHE}) with the mass tensor
of $\widetilde{B}^{\alpha\beta}$.
Then, the following transformation provides another solution:
\begin{equation}
\label{gauge_transf_NG}
q^1 \rightarrow q^1 + c q^I ,
\quad\quad
p_I \rightarrow p_I - c p_1 .
\end{equation}
The proof is given by exactly the same argument done
for Eq. (\ref{gauge_transf}).

This gauge invariant property tells us that we need to fix
the gauge parameter $c$.
For instance, a possible choice could be requiring
$\bar{V}_{,I1}=0$ which was adopted in Ref.~ \cite{HNMM07}.
One can make other choices if they are more convenient  \cite{HNMM07,HNMM08,HNMM09},
and the physical quantities should not depend on this choice.

\subsection{Moving-frame harmonic equation (MFHE)}
\label{sec: MFHE}

Let us summarize the formulation we obtained so far.
The present formulation can be regarded as the harmonic equations
with the {\it moving-frame} Hamiltonian
\begin{equation}
H_M(\xi,\pi)\equiv H(\xi,\pi) - \lambda_I q^I - \lambda_i q^i .
\end{equation}
Equations (\ref{V_dc_NG}) and (\ref{LHE}) can be rewritten as
\begin{eqnarray}
\label{CDFT}
\delta \{ H_M \}_{\pi=0} = 0 ,\\
\label{MFHE}
({\mathcal M}_M)^\alpha_{\beta} f^i_{,\alpha} = (\bar{\mathcal M}_M)^i_j f^j_\beta ,
\quad&&\mbox{or}\quad
({\mathcal M}_M)^\alpha_\beta g^\beta_{,j}
 = (\bar{\mathcal M}_M)^i_j g^\alpha_{,i} .
\end{eqnarray}
Here, the matrix $({\mathcal M}_M)^\alpha_\beta $ is a product of
the mass and potential,
given in the same way as Eq. (\ref{H}) but with $H_M$.
\begin{equation}
B_M^{\alpha\beta}=\widetilde{B}^{\alpha\beta}\equiv \frac{\partial^2 H_M}{\partial\pi_\alpha\partial\pi_\beta} ,\quad
V_M(\xi)\equiv H_M(\xi,\pi=0) .
\end{equation}
It turns out that the LHE becomes identical to
the harmonic equation at the equilibrium with $H_M$.
Therefore, we may call this formulation ``moving-frame harmonic equation''
(MFHE).
It should be noted that the terms $-\lambda_I q^I -\lambda_i q^i$
are not merely the constraints.
These terms changes the mass parameters and the potential.
The theory of the MFHE
is basically equivalent to the gauge-invariant formulation of
the adiabatic self-consistent collective coordinate (ASCC) method \cite{HNMM07}.

From this moving-frame formulation,
it is evident why we use the Hamiltonian in the rotating frame,
$\hat{H}-\lambda \hat{N}$, in Sec.~\ref{sec: FAM_QRPA}.
The same argument is also applicable to 
the quasiparticle random-phase approximation (QRPA)
in the superfluid phase \cite{RS80,BR86}.
Since the ground state does not correspond to
the equilibrium of the energy surface ($dE/dN\neq 0$),
the QRPA is a harmonic approximation at a non-equilibrium state.
According to the present theory, requirements of the covariance and
the extension of the point transformation defines the
{\it moving frame} in which the QRPA should be formulated.

\subsubsection{Practical solution of MFHE}
\label{sec: practical_MFHE}

The theory to define a decoupled submanifold
consists of Eqs. (\ref{CDFT}) and (\ref{MFHE}):
The first equation (\ref{CDFT}) is the potential minimization
with constraints on $q^i$ and $q^I$, which defines the position $\xi$.
The second equation (\ref{MFHE}) defines the normal modes $f^i_{,\alpha}$
($g^\alpha_{,i}$) at the same position $\xi$,
which should provide $f^i_{,\alpha}$ used in Eq. (\ref{CDFT}).
Therefore, these equations should be self-consistently solved.

Let us discuss the $K=1$ case in more details, how to construct the MFHE
matrix $({\mathcal M}_M)^\alpha_\beta$.
The MFHE (\ref{MFHE}) contains higher-order terms which
are not present in the LHE discussed in Sec.~\ref{sec: practical}:
$f^I_{,\alpha\beta}$, $f^{(1)I\alpha\beta}$,
$f^1_{,\alpha\beta}$, and $f^{(1)1\alpha\beta}$.
Among these quantities, 
$f^I_{,\alpha\beta}$ and $f^{(1)I\alpha\beta}$ are calculable
if we know the operators corresponding to $q^I$ explicitly,
such as the particle number and the angular momentum.
The curvature $f^1_{,\alpha\beta}$ can be eliminated by the
same procedure as that in Sec.~\ref{sec: practical}.
Thus, the remaining unknown quantity is $f^{(1)1\alpha\beta}$.

Although we do not have
a general principle to determine $f^{(1)1\alpha\beta}$,
there may be possible prescriptions.
In the case that there is a single constant of motion $q^I$, 
the canonicity condition of Eq.~(\ref{EAT_2}) gives constraints
whose number is same as the number of the index $\alpha$,
namely the number of 2qp states.
Using these constraints,
possible prescriptions are, for instance,
\begin{enumerate}
\item Diagonal assumption:
Assuming $f^{(1)1\alpha\beta}=f^{(1)1\alpha}\delta^{\alpha\beta}$,
$f^{(1)1\alpha}$ can be determined by Eq.~(\ref{EAT_2}).
\item Strong canonicity condition:
Both $q^1$ and $q^I$ are assumed to be represented by
one-body operators $\hat{Q}^1$ and $\hat{Q}^I$, respectively,
where $\hat{Q}^I$ is explicitly known.
Then, requesting $[\hat{Q}^1,\hat{Q}^I]=0$ 
can determine $f^{(1)1\alpha\beta}$.
\end{enumerate}
In the numerical applications in Sec.~\ref{sec: application_LACM},
we adopt the prescription 2 to examine the effect of $f^{(1)1\alpha\beta}$.
Effect of this term turns out to be negligibly small for the multi-O(4)
model \cite{HNMM07}.

After eliminating the curvature terms,
the MFHE can be rewritten in the same form as Eq. (\ref{LHE_1}), with
\begin{equation}
\begin{split}
({\mathcal M}_M)^\alpha_\beta &\equiv \widetilde{B}^{\alpha\gamma}
( V_{,\gamma\beta} - \bar{V}_{,I} f^I_{,\gamma\beta} )
 + \frac{1}{2} \widetilde{B}^{\alpha\gamma}_{,\beta} V_{,\gamma} \\
(\omega_M)^2 &\equiv \bar{B}^{11} \bar{V}_{,11} + \frac{1}{2} \bar{B}^{11}_{,1} \bar{V}_{,1} .
\end{split}
\end{equation}
The equations equivalent to these have been solved in
Refs.  \cite{HNMM07,HNMM08,HNMM09}, with the second prescription given above.

\section{Giant resonances studied with Skyrme EDFs in the linear regime}
\label{sec: GR}

Applications of the TDDFT have been mostly studied
in the linear response regime.
In this section, we show selected results of the applications
of the Green's function method (Sec.~\ref{sec: green_function}) and
the finite amplitude method (Sec.~\ref{sec: FAM}) for nuclei
without the pairing correlations, and
the standard diagonalization method \cite{RS80} for
superfluid nuclei.

\subsection{Giant resonances in the normal phase}
\label{sec: GR_in_NP}

\subsubsection{Coordinate-space representation}
\label{sec: coordinate_space_representation}.

For the Skyrme functionals, which 
is a functional of local one-body densities,
the coordinate-space representation is one of the convenient
choices \cite{FKW78}.
In the followings, we assume $\vec{r}$ involves the spin and isospin
indices, if necessary.
We adopt the three-dimensional (3D) Cartesian grid-space representation
in Sec.~\ref{sec: GFM_application} adn Sec.~\ref{sec: FAM_application}.
Each KS orbital $\phi_i(\vec{r})$ is represented
at discretized grid points $(x_d,y_d,z_d)$.
In the linear regime, behaviors of the TDKS orbitals
$\psi_i(\vec{r},t)$ in the region far outside of the nucleus are
irrelevant in the calculations.
This is because the density response $\rho_1(\vec{r},t)$
vanishes where the KS orbitals in the ground-state $\phi_i(\vec{r})=0$:
\begin{equation}
\rho_1(\vec{r},t)=\sum_i \left\{ |\psi_i(\vec{r},t)|^2 - |\phi_i(\vec{r})|^2
\right\}
=\sum_i \left\{ \psi_{1,i}(\vec{r},t) \phi_i^*(\vec{r}) + \mbox{c.c.}
\right\}
= 0 .
\end{equation}
\begin{wrapfigure}{r}{0.4\textwidth}
\vspace{-10pt}
\includegraphics[width=0.4\textwidth]{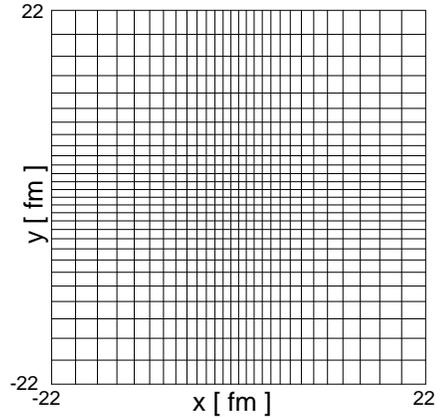}
\caption{\label{fig: grid}
Adaptive grid in the $(x,y)$-plane, used in
calculations in the following sections.
}
\end{wrapfigure}
Thus, we use the 3D grid representation 
with an adaptive mesh \cite{NY05,INY09},
to reduce the number of grid points in the outer region.
See Fig. \ref{fig: grid} for such an example.

The forward and backward amplitudes,
$X_{mi}(\omega)$ and $Y_{mi}(\omega)$, in the linear response
equations (\ref{LRE_FAM}) also possess two indices, $(mi)$.
For these, it is convenient to adopt the mixed representation:
the particle index $m>N$ is replaced by the coordinate
$\vec{r}$, but the hole index $i\leq N$ is kept.
The number of hole (occupied) orbitals in finite nuclei is 
of order of 100, at most.
This mixed representation is adopted in the application of the FAM
in Sec.~\ref{sec: FAM_application}.

\subsubsection{Application of Green's function method}
\label{sec: GFM_application}

We first show applications of the Green's function method.
In the case that the KS orbitals are defined in a potential with
the spherical symmetry $v_s(\vec{r})=v_0(r)$,
this is known under the name of ``continuum RPA'' in the nuclear physics
 \cite{SB75} and has been extensively utilized to study giant resonances
in nuclei \cite{LG76,HSZ98,Sag01}.
It should be also noted that the extension to the linear density response in
superfluid systems has been achieved with the use of the
anomalous Green's function  \cite{Mat01}.

For deformed systems, the construction of the Green's function with
a proper boundary condition involves a significant task \cite{LS84,NY01}
and the applications to nuclear systems are still very limited.
We adopt an approach using a double iterative algorithm \cite{NY01,NY05}.
Roughly speaking, this is based on the fact that
Eqs. (\ref{rho_1_scf_FT}) and (\ref{v_scf}) are
rewritten in a form of the linear algebraic equation
with respect to $\rho_1(\omega)$,
In addition, the action of the Green's function,
$\ket{\psi^{(\pm)}}=G_s^{(\pm)}(E) \ket{\phi}$ for a given state
$\ket{\phi}$ is also given by a solution of the
linear equation.
We solve these linear algebraic equations by using the iterative
methods.
See Refs.~ \cite{NY01} for details.

\begin{figure}[tb]
\centerline{
\includegraphics[width=0.5\textwidth]{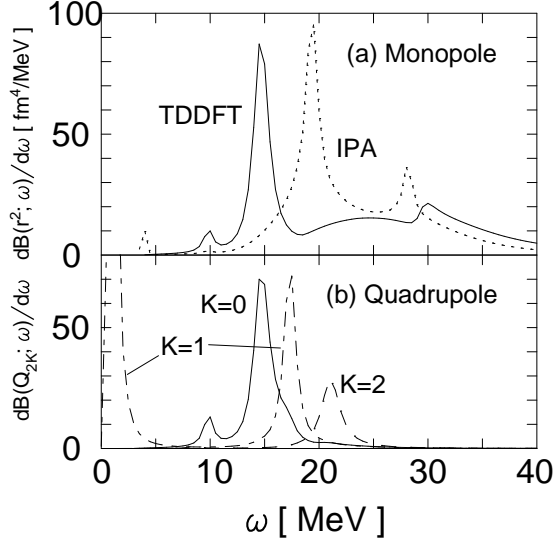}
}
\caption{\label{fig: Ne20_GFM}
Strength functions for $^{20}$Ne calculated for
complex frequencies $\omega+i\gamma/2$ with
the smoothing parameter $\gamma=1$ MeV \cite{NY05}.
(a) isoscalar monopole strengths.
The solid line corresponds to the full response (``TDDFT''),
while the dotted line is obtained by neglecting the residual kernel, that is
indicated by ``IPA'' (independent particle approximation).
(b) isoscalar quadrupole strengths.
The $K=0$, $K=1$, and $K=2$ quadrupole strengths are
shown by solid, dash-dotted, and dashed lines, respectively.
}
\end{figure}
In Fig.~\ref{fig: Ne20_GFM}, we demonstrate an example of the results of the
present iterative algorithms for deformed systems.
The isoscalar monopole and quadrupole strength functions in $^{20}$Ne
are calculated with the BKN energy functional
that is a simplified version of the Skyrme functional \cite{BKN76}.
In nuclear binding energy,
there is a strong cancellation
between the positive kinetic energy and the negative potential energy.
The large nucleonic kinetic energy plays an important role in many phenomena
in nuclei.
The giant quadrupole resonance (GQR) is such an example.
Namely, the restoring force for the vibrational motion mainly comes from the
distortion of the Fermi sphere in the momentum space  \cite{Bla80}.

The GQR shows three peaks in order of
$K=0$, 1, and 2 in increasing energy (Fig.~\ref{fig: Ne20_GFM} (b)).
This is because the ground state has a superdeformed prolate shape with 
$\beta\approx 0.6$.
The result also indicates no low-energy quadrupole vibration
except for the NG mode with $K=1$.
This is a characteristic feature of the superdeformation  \cite{NMM92,NMMS96}.

The monopole strength consists of two components:
a peak at 15 MeV and a broad hump in the energy region of $E>20$ MeV.
The dotted line indicates the strength of the independent particles
obtained by $\Pi_s$.
The residual kernel $w(\omega)$ shifts
the two components to opposite directions.
The peak at $\omega\approx 20$ MeV is shifted to lower energy
by about 5 MeV.
This lowering in energy is due to
strong coupling to the quadrupole resonance.
In fact, the peak lies at exactly the same energy as the $K=0$ quadrupole
resonance (Fig.~\ref{fig: Ne20_GFM}~(b)).

The calculated single-particle energy of the last occupied orbital
is $-10.8$ MeV.
Thus, all the high-energy peaks in Fig.~\ref{fig: Ne20_GFM}
are embedded in the continuum.
The broad structure of the monopole strength function at $E>20$ MeV
indicates that there is no prominent monopole resonance in this nucleus,
except for the peak due to the coupling to the GQR.

\subsubsection{Application of FAM}
\label{sec: FAM_application}

The FAM is a feasible approach to the linear response calculations
with realistic EDF.
With a Skyrme-type EDF,
the FAM formula (\ref{FAM_2}) tells us to calculate
the operation of $v_{\rm ks,1}(\vec{r},\omega)$ in the
coordinate space as
\begin{equation}
v_{\rm ks,1}(\vec{r},\omega)\phi_i(\vec{r}) = \frac{1}{\eta} 
\left(
h\big[\bar\psi_\eta',\bar\psi_\eta^* \big](\vec{r})\phi_i(\vec{r})
    - \epsilon_i\phi_i(\vec{r}) \right),
\end{equation}
with $\bar\psi_{\eta,i}^*(\vec{r})=\phi_i^*(\vec{r})+\eta Y_i^*(\omega,\vec{r})$ and
$\bar\psi'_{\eta,i}(\vec{r})=\phi_i(\vec{r})+\eta X_i(\vec{r},\omega)$.
Exchanging the forward and backward amplitudes
in $\bar\psi_{\eta,i}(\vec{r})$ and $\bar\psi'_{\eta,i}(\vec{r})$,
we may calculate
$v_{ks,1}^\dagger(\vec{r},\omega)\phi_i(\vec{r})$ in the same way.
Adopting the local external field
$
v_1(\vec{r},\omega)= F(\vec{r})
$,
the strength function is calculated from the
obtained forward and backward amplitudes,
as in Eqs.~(\ref{S_F_2}) and (\ref{R_F_2}),
\begin{eqnarray}
S_F(\omega)=
 -\frac{1}{\pi} \mbox{Im} \sum_i\int d\vec{r} \left\{
      \phi_i^*(\vec{r}) F^\dagger(\vec{r}) X_i(\vec{r},\omega)
      + Y_i^*(\vec{r},\omega) F^\dagger(\vec{r}) \phi_i(\vec{r}) \right\} .
\end{eqnarray}

The FAM makes a coding of the linear response calculation
much easier than the other methods.
The FAM does not require explicit construction
of the matrix, thus, it significantly reduces a memory resource requirement.
These are the main advantages of the FAM.
In addition, 
the computational task scales linearly both
with the size of the model space and with the particle number.
This linear dependence was confirmed in the actual calculations as well.
Therefore, the FAM may demonstrate its merit for larger systems.

A disadvantage is the fact that the iterative procedure is difficult
to parallelize.
Since the calculations with different $\omega$ are independent,
this provides a trivial parallelization with respect to $\omega$.
This leads to a use of PC cluster systems with
128-256 processors in parallel,

\medskip
\noindent
\underline{Choice of iterative algorithms}
\begin{figure}[tb]
\centering
\includegraphics[width=0.8\textwidth]{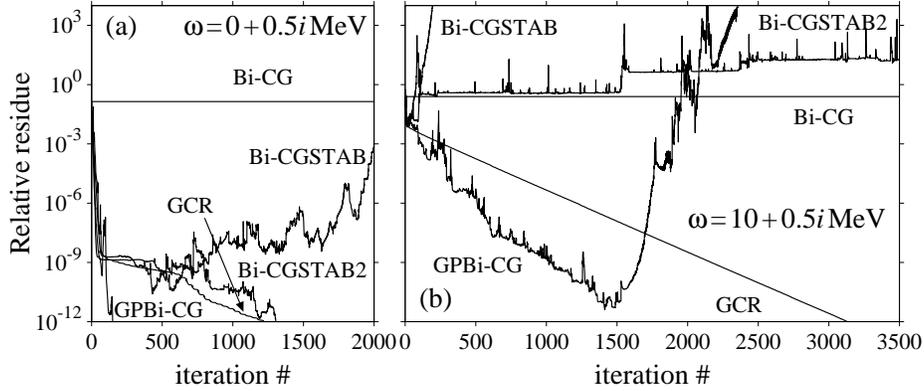}
\caption{
Convergence property of different iterative methods:
Relative residue, $r_n$,
is shown as a function of iteration number $n$,
at complex frequencies of
$\omega=0+0.5i$ MeV (a) and $10+0.5i$ MeV (b).
See text for details.
The figure is taken from Ref.  \cite{INY09}.
}
\label{fig: convergence_FAM}
\end{figure}

To solve the linear response equations~(\ref{LRE_FAM}),
an iterative method is utilized.
Here, we denote this equation symbolically as $\mathbf{A} \vec{x} = \vec{b}$.
For the Skyrme energy functional, the matrix $\mathbf{A}$
in the $\vec{r}$-space grid representation is sparse.
Therefore, the iterative methods, such as the conjugate gradient
(CG) method  \cite{PTVF07}, should work efficiently.
However, since we calculate for the complex frequency $\omega$,
the matrix $\mathbf{A}$ is not Hermitian.
Therefore, we should adopt one of a number of variants of the CG method
extended for non-hermitian problems.
In Fig.~\ref{fig: convergence_FAM},
we show performance of some of different iterative algorithms:
Bi-conjugate gradient (Bi-CG) method
 \cite{PTVF07}, generalized conjugate residual (GCR) method
 \cite{EES83}, generalized product-type bi-conjugate gradient
(GPBi-CG) method  \cite{Zhang97}, Bi-CGSTAB method \cite{Vorst92},
and Bi-CGSTAB2 method  \cite{Gutk93}.
The magnitude of the relative residue,
\begin{equation}
r_n = \vert \vec{b}-\mathbf{A}\vec{x_n} \vert 
/ \vert \vec{b} \vert
\label{residue}
\end{equation}
is plotted against the number of iterations,
for the case of the electric dipole response in $^{16}$O,
The initial vector is taken as $\vec{x}=0$.

It turns out the convergence property depends on the frequency.
At low frequency ($\omega=0+0.5i$ MeV),
all the solvers except for the Bi-CG method quickly reach the convergence.
On the other hand, at higher frequency ($\omega=10+0.5i$ MeV),
only the GCR and the GPBi-CG methods lead to the convergence.
In Fig.~\ref{fig: convergence_FAM},
the GCR shows the most stable behavior for the convergence,
though it requires larger computer memory resources
than other methods.
Recently, we have also tested the generalized product-type bi-conjugate
gradient method with associated residual (GPBiCG-AR)  \cite{TF08},
which indicates a better performance.
It should be noted that we need much smaller number of iteration
to reach the convergence in the harmonic-oscillator-basis representation
 \cite{Sto11}.
The coordinate space of a relatively large 3D box size contains
a large number of irrelevant mesh points, which perhaps makes
the convergence very slow.
It should be noted that an iterative algorithm based on the
Arnoldi diagonalization method was proposed for similar problems  \cite{Toi10}.

\medskip
\noindent
\underline{Nuclear photoabsorption cross sections}
\begin{figure}[tb]
\begin{center}
\includegraphics[width=0.8\textwidth,keepaspectratio]{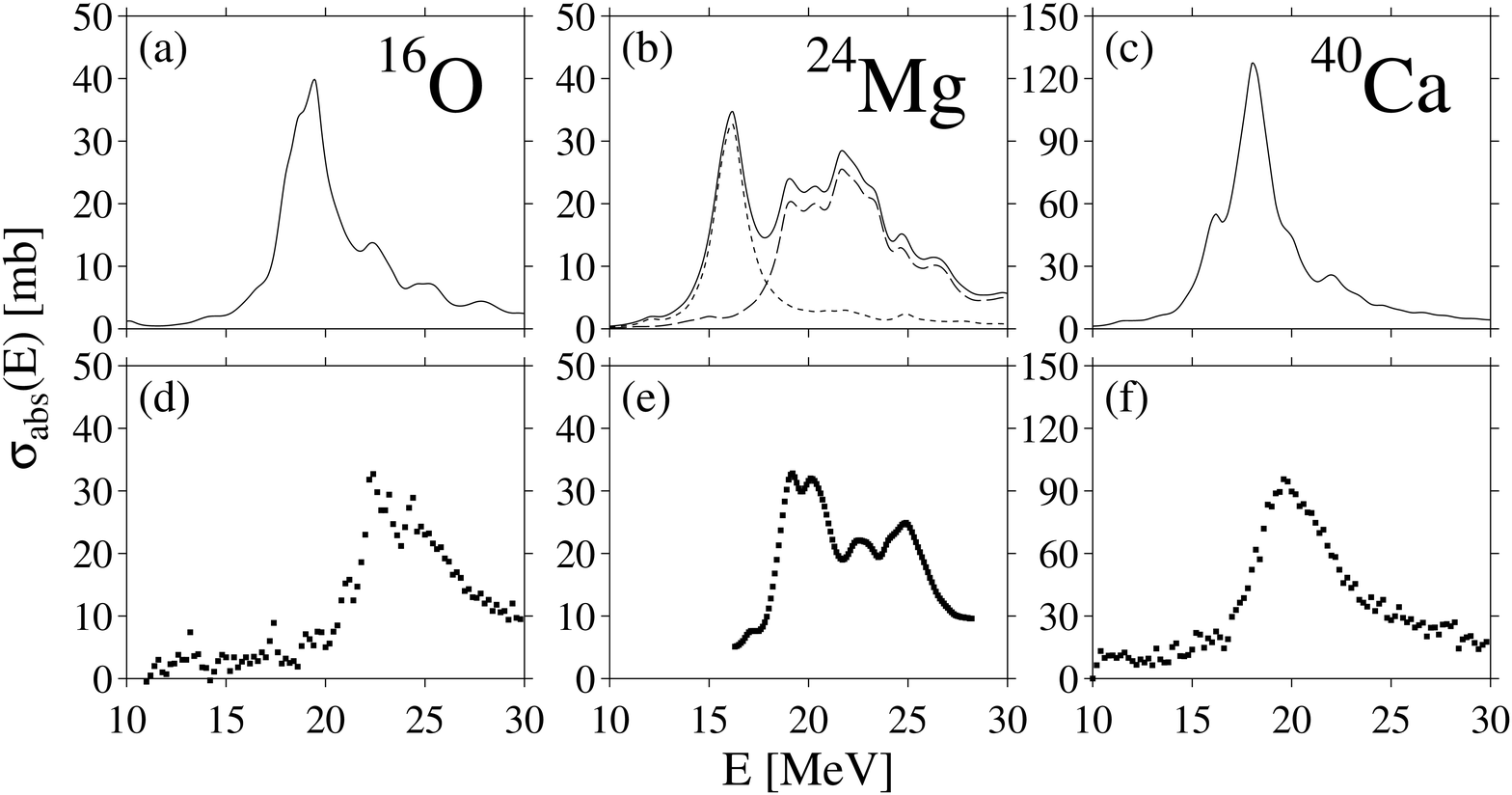}
\caption{{\small
Calculated (a-c)  \cite{INY09} and experimental (d-f)
 \cite{Ishkhanov02,Varlamov03} photoabsorption cross sections in
$^{16}$O, $^{40}$Ca, and $^{24}$Mg.
We use
the SkM$^\ast$ parameter set and $\gamma=1$ MeV.
}}
\label{fig: 16O.24Mg.40Ca}
\end{center}
\end{figure}

Adopting the electric dipole ($E1$) operator with the $E1$ recoil charges
as the operator $F$,
we calculate the $E1$ strength function $S_{E1}(E)$ that is
converted into the photoabsorption cross section  $\sigma_{\rm abs}(E)$
in the dipole approximation.
Calculated photoabsorption cross sections for spherical nuclei,
$^{16}$O, $^{40}$Ca and deformed nucleus $^{24}$Mg,
are compared with experimental data
in Fig. \ref{fig: 16O.24Mg.40Ca}.
Here, the complex frequency $\omega+i\gamma/2$ with
the width $\gamma=1$ MeV is adopted.
In each nucleus, there is a broad peak in $\sigma_{\rm abs}(E)$
around $E=20-25$ MeV, which corresponds to the giant dipole resonance (GDR).
The overall profile of the experimental cross section is well reproduced,
though the calculated energies of the GDR
peaks are underestimated by a few MeV.
The discrepancy is more prominent for lighter nuclei,
which is observed with almost
all the Skyrme energy functionals  \cite{NY05,INY09}.

For spherical nuclei, the
GDR widths calculated with $\gamma=1$ MeV are narrower than
the corresponding experimental data.
This seems to suggest that
the spreading width $\Gamma^\downarrow$,
which takes account of effects decaying into compound states,
such as two-particle-two-hole excitations,
is slightly larger than $\gamma=1$ MeV.
For the deformed nucleus $^{24}$Mg, the GDR
peak splitting caused by the ground-state deformation well agree with
the experiments, although the magnitude of the deformation splitting is
slightly too large in the calculation.
We may interpret that the experimental GDR
peak around $E=20$ MeV is associated with the $K=0$ mode,
and those at $E=22\sim 25$ MeV correspond to the $K=1$ mode.
The double-peak structure of the $K=1$ GDR peak is well reproduced.
Approximately, the calculated cross section is shifted to lower energy
from the experimental ones, by about 3 MeV.

\begin{figure}[tb]
\begin{center}
\includegraphics[width=0.8\textwidth,keepaspectratio]{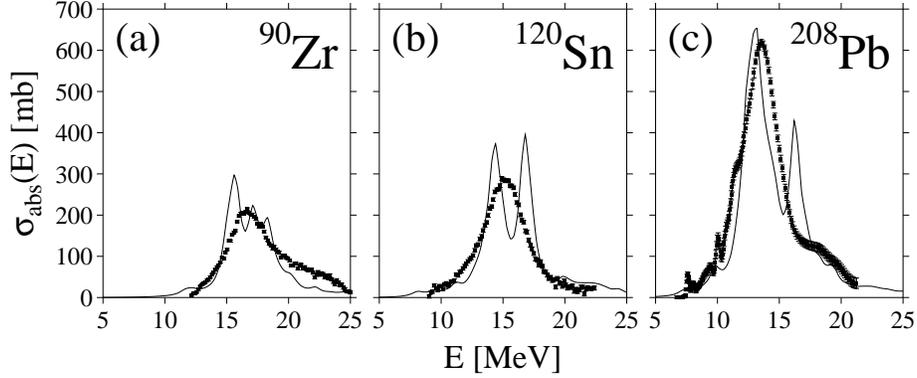}
\caption{{\small
Calculated photoabsorption cross sections
for (a) $^{90}$Zr, (b) $^{120}$Sn, and (c) $^{208}$Pb.
The calculation has been performed with 
the SkM$^*$ parameter set, and $\gamma=1$ MeV  \cite{INY09}.
The experimental data (symbols) are taken from
Refs.  \cite{Lepretre71, Lepretre74, Beljaev91}.
}}
\label{fig.comp.exp.heavy}
\end{center}
\end{figure}

For heavier nuclei, the calculation better agrees with experiments \cite{INY09}.
Calculated photoabsorption cross sections
in spherical nuclei $^{90}$Zr, $^{120}$Sn, and $^{208}$Pb 
are compared with experimental data
in Fig. \ref{fig.comp.exp.heavy}.
The calculated GDR peak shows a splitting, however, this may be due to
the spurious effect coming from the box discretization.
Except for this splitting, the results agree well with the experimental data.
A single Lorentzian fit for the
photoabsorption cross section gives
the GDR peak energies of 16.4, 15.2, and 13.3 MeV
for $^{90}$Zr, $^{120}$Sn, and $^{208}$Pb, respectively.
The corresponding experimental values are
16.7, 15.4, 13.6 MeV, respectively.
The GDR peak positions are well reproduced
within an error of 400 keV.
We may conclude that the SkM$^*$ functional
reproduces peak energies of the $E1$ resonances in heavy nuclei.

For heavy nuclei, the spreading width was supposed to be a major
part of the total damping width  \cite{Wam88,HW01}.
However,
the artificial width of $\gamma=1$ MeV, which is supposed to take account
of missing spreading effects, reproduces
the observed GDR width.
Although the total damping width is about 4 MeV for these nuclei,
the spreading width is less than half of the total width.
In fact, the fragmentation of the strength into non-collective 1p-1h
states (Landau damping)
is significant in the present calculation.
Thus, the small spreading width (about 1 MeV) is able to
reproduce a broadening of the experimental strength distribution.
This is also consistent with other recent calculations  \cite{Ter06,Sil06}.

\subsection{Giant resonances in the superfluid phase}
\label{GR_in_SP}

Inclusion of the pair density for systems with superfluidity is,
theoretically, a straightforward extension of the TDKS to
TDKSB equation (See Sec.~\ref{sec: TDKSB}).
However, in practice, it costs a significant increase in numerical task.
For instance, the number of matrix elements in
Eq. (\ref{RPA_eigenvalue}) is roughly proportional to
$M^2$ for the normal system and $M^4$ for superfluid systems,
where $M$ is a dimension of the single-particle model space.
At present, it is difficult to adopt
the 3D coordinate-space representation in Sec.~\ref{sec: GR_in_NP}
for superfluid nuclei \cite{SBMR11}.
In this section, we adopt the symmetry restrictions on the shape of
the potentials to reduce the numerical costs.

\begin{figure}[t]
\begin{center}
\includegraphics[scale=0.66]{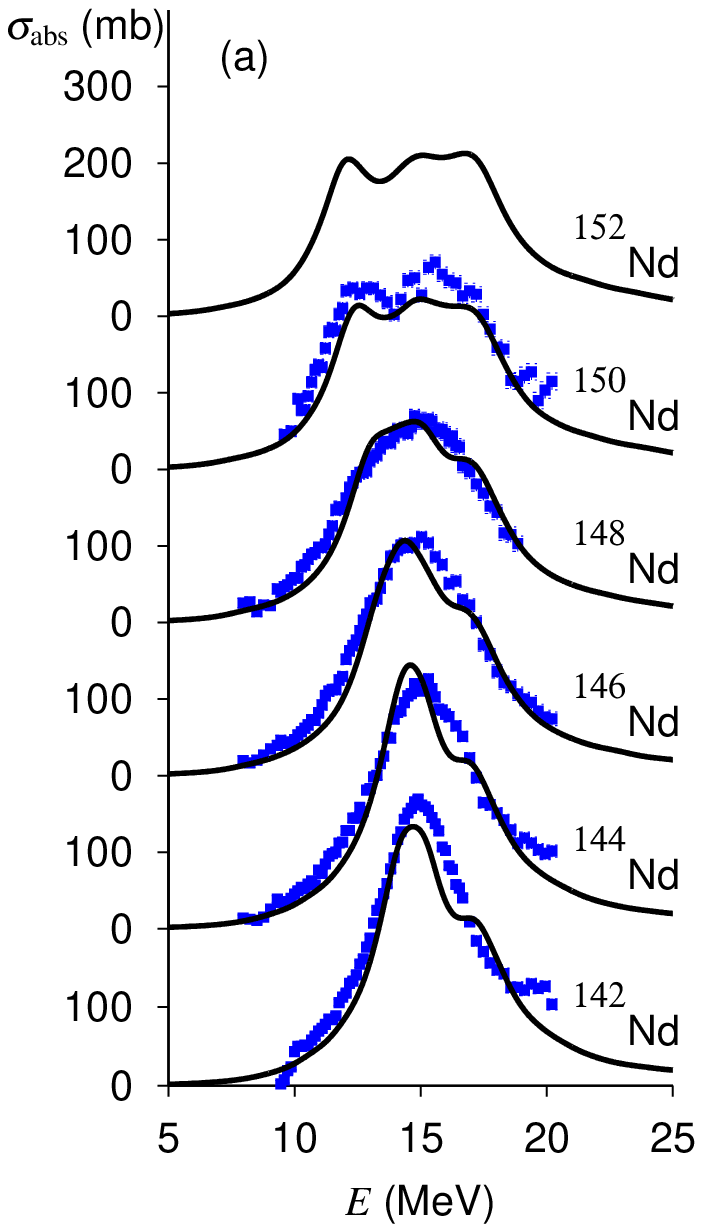}
\includegraphics[scale=0.66]{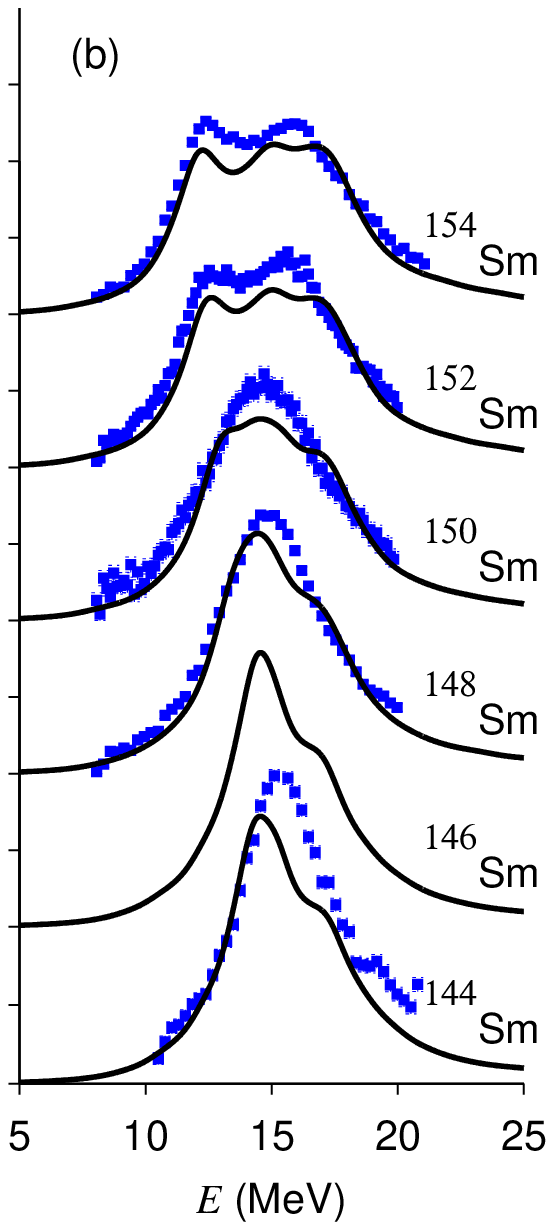}
\caption{
Photoabsorption cross sections in (a) Nd and (b) Sm isotopes as functions of
photon energy \cite{YN11}.
The Skyrme energy functional with the SkM* parameter set was used.
The experimental data~ \cite{car71,car74} are denoted by filled squares.
}
\label{fig: Nd_Sm}
\end{center}
\end{figure}
In the linear response equation (\ref{RPA}),
the system can oscillate
without the external perturbation $v_1(\omega)=0$ (${\mathcal V}_1(\omega)=0$),
at the eigenfrequencies $\omega=\omega_n$,
Thus, the normal modes of excitation are
obtained by solving the eigenvalue equation:
\begin{equation}
\label{RPA_eigenvalue}
\begin{pmatrix}
A   & B \\
B^* & A^*
\end{pmatrix}
\begin{pmatrix}
X^n \\
Y^n
\end{pmatrix}
=
\omega_n
\begin{pmatrix}
1 & 0 \\
0 & -1
\end{pmatrix}
\begin{pmatrix}
X^n \\
Y^n
\end{pmatrix}
.
\end{equation}
Solving this eigenvalue equation in the
qp basis is the most common method for TDDFT
to study elementary modes of excitation in superfluid nuclei.
Because of numerical difficulties, most of current studies
with Skyrme EDFs for deformed nuclei are restricted to axially deformed nuclei
 \cite{LPDVB10,Sto11,YN11,hfbtho}.

We first calculate the qp states in
the ground state in the 2D coordinate space assuming the axial symmetry.
Then, all the necessary quantities are expressed
in the qp representation.
For deformed systems, the number of 2qp states becomes huge
and we often need a further truncation of the 2qp space  \cite{YN11}.
In addition, the residual kernel associated with the
long-range Coulomb part is neglected in the present calculation.
The solutions of the eigenvalue equation (\ref{RPA_eigenvalue}) are
obtained using the symmetrization procedure \cite{RS80}.

Heavy nuclei with open-shell configurations are supposed to have
a superfluid character caused by the neutron-neutron and proton-proton
pairing correlations.
Here, we show results for Nd and Sm isotopes.
The protons have finite pair densities for all these isotopes, while
the neutron pair density vanishes for $^{142}$Nd and $^{144}$Sm which
correspond to the neutron magic number $N=82$.
These isotopes exhibit typical examples of the quantum phase transitions
in their ground states,
from spherical to prolate shapes, and simultaneously,
from the normal to superfluid phases,
as increasing the neutron number.
Actually, the calculated ground states show a spherical shape for
$^{142,144}$Nd and $^{144,146}$Sm, and a prolate shape for the others.
The magnitude of deformation increases
as the neutron number changes from 86 to 92.

In Fig.~\ref{fig: Nd_Sm}, the calculated photoabsorption cross sections
for Nd and Sm isotopes are shown
together with the available experimental data~ \cite{car71,car74}.
The dipole strength at discrete eigenenergies are smeared by the
Lorentzian with a width $\gamma=2$ MeV.
The GDR peak energies well agree with experimental values, and
produces the deformation splitting in $^{150,152}$Nd and $^{152,154}$Sm.
The GDR width calculated with $\gamma=2$ MeV
is also in good accordance with the experimental values.
The nice agreement on the broadening indicates that
the smearing width $\gamma=2$ MeV
has a good correspondence with the spreading width
$\Gamma^{\downarrow}$ in these nuclei.

The isotopic dependence of the peak broadening
is well reproduced, even for the transitional nuclei.
The width for $N=82$ and 84 is calculated as $\Gamma\approx 4.5$ MeV,
and it gradually increases to about 6 MeV for $N=88$
($^{148}$Nd and $^{150}$Sm), then
the peak splitting becomes visible for $N\geq 90$ and 92.
Here, the width $\Gamma$ is evaluated by fitting the calculated cross section
with a Lorentz line.
The broadening of the GDR was found to be well correlated with the
nuclear quadrupole moments \cite{Oka55,Dan58}.
Thus, it is interpreted as the mode-mode coupling effects to the low-lying
collective modes~ \cite{HW01}.
In the present calculation,
the mode coupling is not explicitly taken into account.
However, the linear response based on the deformed state may implicitly
include a part of the coupling effect.
Figure~\ref{fig: Nd_Sm} shows that the isotopic dependence
can be well accounted for
by the gradual increase of the ground-state deformation.

\section{Large amplitude collective dynamics in shape coexistence}
\label{sec: application_LACM}

Low-energy collective modes of excitation in nuclei present
unique features of the finite quantum systems.
In contrast to giant resonances discussed in Sec.~\ref{sec: GR},
the linear approximation is often insufficient for low-lying
collective states in nuclei.
The vibrational excitations should contain a strong anharmonicity
when the stability matrix ${\mathcal S}$
has an eigenvalue close to zero, namely when the system
is close to the critical point of the stability.
This kind of situation occurs in many nuclei, especially for the
quadrupole modes of excitation in a transitional situation
such as shape phase transition (Fig.~\ref{fig: Nd_Sm}) and
shape coexistence phenomena \cite{Woo92,HW11}.
Therefore, we need to go beyond the linear regime
for describing these nuclear
phenomena of a large amplitude nature.

One of the unique features of the low-lying collective motion
in nuclei is the fact that its character
significantly changes from nucleus to nucleus.
In addition, its structure is affected by an interplay
between the pairing and deformation correlations, which
often results in a spontaneous breaking of symmetry.
Therefore, it is difficult to introduce an {\it a priori} assumption
on the nature of the low-lying collective motion.
Therefore, the theory presented in Sec.~\ref{sec: LACM}
is suitable to find optimal collective manifold
which leads to a collective Hamiltonian.
In this section, we show applications of the theory in Sec.~\ref{sec: LACM}.

In this section, we apply the method in Sec.~\ref{sec: MFHE}
to description of low-lying spectra
in neutron-deficient $^{68}$Se.
This nucleus shows a feature of the shape coexistence:
The experimental data indicate that there exist rotational bands with
different characters,
which has been interpreted to be those with prolate and oblate shapes
 \cite{Fischer00,Fischer03}.
The mixing property of the two bands are of significant interest,
which influences the excitation spectra and transition probabilities.
In this section, we show results only for $^{68}$Se.
The same analysis on $^{70,72}$Se can be found in Ref.  \cite{HNMM09}.

\subsubsection{Pairing-plus-quadrupole model}

The pairing-plus-quadrupole (P+Q) model is one of the most successful
models that allows us to describe nuclear phenomena involving
the quadrupole and pairing degrees of freedom.
Baranger and Kumar studied the quadrupole motion in the P+Q model,
assuming that the collective coordinates are given by the quadrupole
deformations $(\beta,\gamma)$, and that the collective mass parameters
are given by the cranking formula with phenomenological corrections
 \cite{BK65}.
However, a study of the same model \cite{NWD99} reveals that, even at
the minimum point of the potential, the self-consistent mass parameters
and the property of normal modes are very different from those
utilized by Baranger and Kumar.
In the followings, we also adopt the Hamiltonian similar to the P+Q model
and study the large amplitude collective motion in Se nuclei.

The P+Q model in the present study includes the quadrupole pairing and
is given by
\begin{equation}
\begin{split}
\hat{H}=&\sum_k \epsilon_k c_k^\dag c_k
-\frac{1}{2}\sum_{\tau=n,p} G_0^{(\tau)} 
 (\hat{A}^{(\tau)\dag} \hat{A}^{(\tau)} + \hat{A}^{(\tau)}\hat{A}^{(\tau)\dag})
 \\
&  - \frac{1}{2} \sum_{\tau=n,p}G_2^{(\tau)} \sum_K (\hat{B}_{2K}^{(\tau)\dag} \hat{B}_{2K}^{(\tau)} + \hat{B}_{2K}^{(\tau)}\hat{B}_{2K}^{(\tau)\dagger})
 - \frac{1}{2} \chi \sum_K \hat{D}_{2K}^\dag \hat{D}_{2K} ,
\end{split}
\end{equation}
where $\epsilon_k$ are the spherical single-particle energies and
the index $k$ denotes the set of quantum numbers $(njlm)$.
The monopole pairing operator $\hat{A}^{(\tau)\dag}$,
the quadrupole pairing operator $\hat{B}^{(\tau)\dag}$,
and the mass quadrupole operator $\hat{D}^{\dag}$ are defined by
\begin{equation}
\begin{split}
\hat{A}^{(\tau)\dag} =&
 \sum_{(k>0)\in\tau} c_k^{\dag} c_{\bar{k}}^{\dag}, \quad\quad
\hat{B}_{2K}^{(\tau)\dag} = \sum_{k,l\in\tau} D_{2K}(kl)
c_k^{\dag} c_l^{\dag},\\
\hat{D}_{2K} =& \sum_{k,l} D_{2K}(kl) c_k^\dag c_l ,
\end{split}
\end{equation}
where $D_{2K}(kl)$ are the (dimensionless) quadrupole matrix elements,
modified according to a prescription given by Baranger and Kumar  \cite{BK65}.
The model parameters are determined by adjusting the potential energy surface
calculated with the Skyrme energy functional.
Those values are found in Ref.~ \cite{HNMM09}.

\subsubsection{Results of MFHE}

\begin{figure}[tb]
\begin{center}
\includegraphics[width=0.7\textwidth]{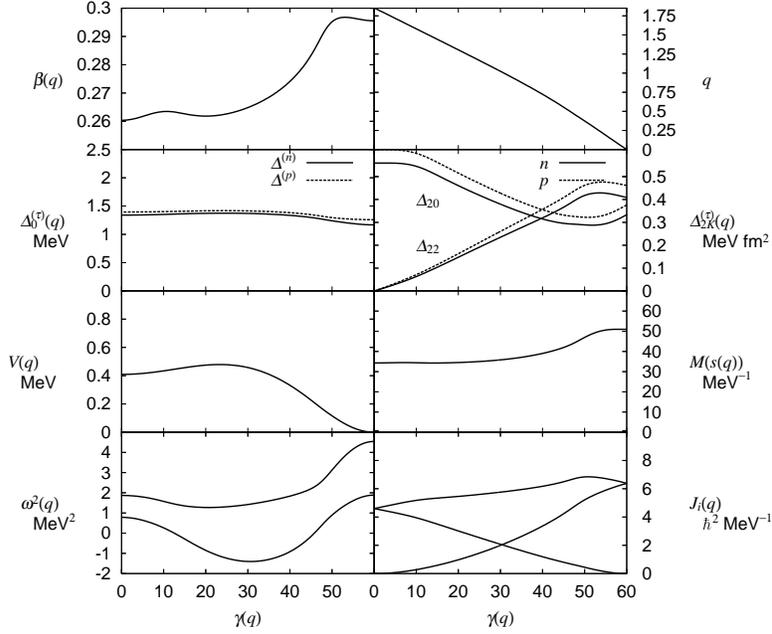}
\end{center}
\caption{
 Calculated quantities on the collective path for $^{68}$Se
 as functions of the triaxiality $\gamma$ \cite{HNMM09}.
 The axial deformation $\beta(q)$,
 the canonical collective coordinate $q$,
 the monopole pairing gaps $\Delta_0^{(\tau)}(q)$,
 the quadrupole pairing gaps $\Delta^{(\tau)}_{20}(q)$ and
 $\Delta^{(\tau)}_{22}(q)$,
 the collective potential $V(q)$, the collective mass $M(s(q))$,
 the lowest two eigenfrequencies of the MFHE
 $\omega^2(q)$, and
 the rotational moments of inertia ${\mathcal J}_i(q)$.
}
\label{fig: 68Se_ASCC}
\end{figure}
The Hartree-Bogoliubov (HB) approximation
to the present P+Q Hamiltonian leads to
two minima (HB states) in the potential energy surface,
in $^{68}$Se:
The lowest minimum has an oblate shape ($\beta=0.3$) and the other
minimum with a prolate shape ($\beta=0.26$)
lies at energy about 400 keV higher than the oblate one.
The obtained collective coordinate $q^1$ has a good
one-to-one correspondence to the triaxial deformation parameter $\gamma$
in the quadrupole deformations $(\beta,\gamma)$ defined by
\begin{equation}
\beta \cos\gamma =  \bra{\Psi(q^1)} \hat{D}_{20} \ket{\Psi(q^1)} ,
\quad
\beta \sin\gamma =  \bra{\Psi(q^1)} (\hat{D}_{22}+ \hat{D}_{2-2}) \ket{\Psi(q^1)}/\sqrt{2} .
\end{equation}
This is seen in the top-right panel in Fig. \ref{fig: 68Se_ASCC}.
In Fig. \ref{fig: 68Se_ASCC}, a variety of quantities calculated on the
collective path are plotted as functions of $\gamma$.
The mass parameter is defined with respect to the geometrical length $s$
in the $(\beta,\gamma)$ plane,
$ds^2=\sqrt{d\beta^2+\beta^2 d\gamma^2}$.
\begin{equation}
M_s^{-1}(q^1)=\left(\frac{ds}{dq^1}\right)^2 \bar{B}^{11}
=\left\{
\left(\frac{d\beta}{dq^1}\right)^2 + \beta^2 \left(\frac{d\gamma}{dq^1}\right)^2
\right\}
\bar{B}^{qq} .
\end{equation}
In the prolate and oblate minima, the lowest modes of excitation in
the MFHE correspond to the gamma vibrations, and the second lowest
to the beta vibrations.
The frequency $\omega$, then, turns into imaginary in the triaxial
region ($10^\circ < \gamma < 50^\circ$).
The imaginary frequency causes no problem in the solution of the MFHE.
We also calculate the moments of inertia by solving the Thouless-Valatin
equations \cite{TV62} on the collective path, which gives
${\mathcal J}_k(q)$, $k=x,y,z$.
These are also shown in the bottom-right panel in Fig.~\ref{fig: 68Se_ASCC}.
They have a characteristic feature similar to the moments
of inertia of the irrotational fluid \cite{RS80}.

\subsubsection{Collective Hamiltonian}

In the total kinetic energy, the position-dependent rotational energy
is added to the one of the one-dimensional shape vibration described
by the coordinate $q^1$.
\begin{equation}
 T = \frac{1}{2}\bar{B}^{ij}(q^1)p_i p_j
\end{equation}
where $i,j=1,\cdots,4$.
Here, $(p_2,p_3,p_4)$ are the total angular momentum $(I_x,I_y,I_z)$,
and $\bar{B}^{ij}=\delta^{ij}
 (\bar{B}^{11}, {\mathcal J}_x^{-1}, {\mathcal J}_y^{-1}, {\mathcal J}_z^{-1})$.
The kinetic energy term is requantized by means of the Pauli prescription:
\begin{align}
 \hat{T} =& - \frac{1}{2}\sum_{ij} |G(q^1)|^{-\frac{1}{2}}
 \frac{\partial}{\partial q^i}
 |G(q^1)|^{\frac{1}{2}} \bar{B}^{ij}(q^1)) \frac{\partial}{\partial q^j} \nonumber \\
 =& - \frac{1}{2}\frac{\partial}{\partial q^1} \bar{B}^{11}(q^1) \frac{\partial}{\partial q^1} 
- \frac{1}{4} \frac{\partial G}{\partial q^1} 
  \frac{\bar{B}^{11}(q^1)}{G(q^1)} \frac{\partial}{\partial q^1} + 
 \sum_{k=x,y,z} \frac{\hat{I}^2_k}{2{\mathcal J}_k(q^1)}, \label{eq:req-kinetic}
 \end{align}
where $G(q)=\bar{B}_{11}(q^1){\mathcal J}_x(q^1){\mathcal J}_y(q^1){\mathcal J}_z(q^1)$ is
the determinant of the metric $\bar{B}_{ij}(q^1)$.
The three components $\hat{I}_k$ of the angular momentum operator are
defined with respect to the principal axes $(x,y,z)$
associated with the intrinsic (moving-frame) state $\ket{\phi(q^1)}$.

The collective Schr\"odinger equation is thus given, with $q^1$
replaced by $q$ hereafter, as
\begin{align}
 ( \hat{T} + V(q) ) \Psi_{IMn}(q,\Omega) = E_{I,n} \Psi_{IMn}(q,\Omega),
 \label{eq:Schroedinger}
\end{align}
where $\Psi_{IMn}(q,\Omega)$ represents the collective wave function
in the laboratory frame.
It is a function of the collective coordinate $q$ and
the three Euler angles $\Omega$,
and specified by the total angular momentum $I$, its projection $M$
on the laboratory $z$-axis,
and the index $n$ distinguishing different quantum states having
the same $I$ and $M$.
Using the rotational wave functions ${\mathcal D}^I_{MK}(\Omega)$,
the collective wave functions in the laboratory frame is written as
\begin{align}
 \Psi_{IMn}(q,\Omega) =& \sum_{K=0}^I \Phi_{IKn}(q) \langle\Omega|IMK\rangle, \label{eq:collwave} \\
 \langle\Omega|IMK\rangle =&
 \sqrt{\frac{2I+1}{16\pi^2(1+\delta_{K0})}}(
 {\mathcal D}^I_{MK}(\Omega) + (-)^I {\mathcal D}^I_{M-K}(\Omega))
\end{align}
where the sum in Eq.~(\ref{eq:collwave}) is restricted to even $K$.
Here, $\Phi_{IKn}(q)$ represents the shape vibrational motion
described by the coordinate $q$.

Normalization of the vibrational part of the collective wave functions
is given by
\begin{align}
 \int d\tau' \sum_{K=0}^I \Phi^\ast_{IKn}(q) \Phi_{IKn'}(q) = \delta_{nn'}
\end{align}
where the volume element is
\begin{align}
 d\tau = d\tau' d\Omega = \sqrt{|G(q)|}dq d\Omega .
 \label{eq:metric}
\end{align}
The boundary conditions for the collective Schr\"odinger equation
(\ref{eq:Schroedinger}) can be specified by projecting the obtained collective
path onto the $(\beta,\gamma)$ plane
and by using the symmetry properties of the Bohr-Mottelson
collective Hamiltonian  \cite{KB67}.

\begin{figure}[tb]
\begin{center}
\begin{tabular}{cc}
\includegraphics[width=0.5\textwidth]{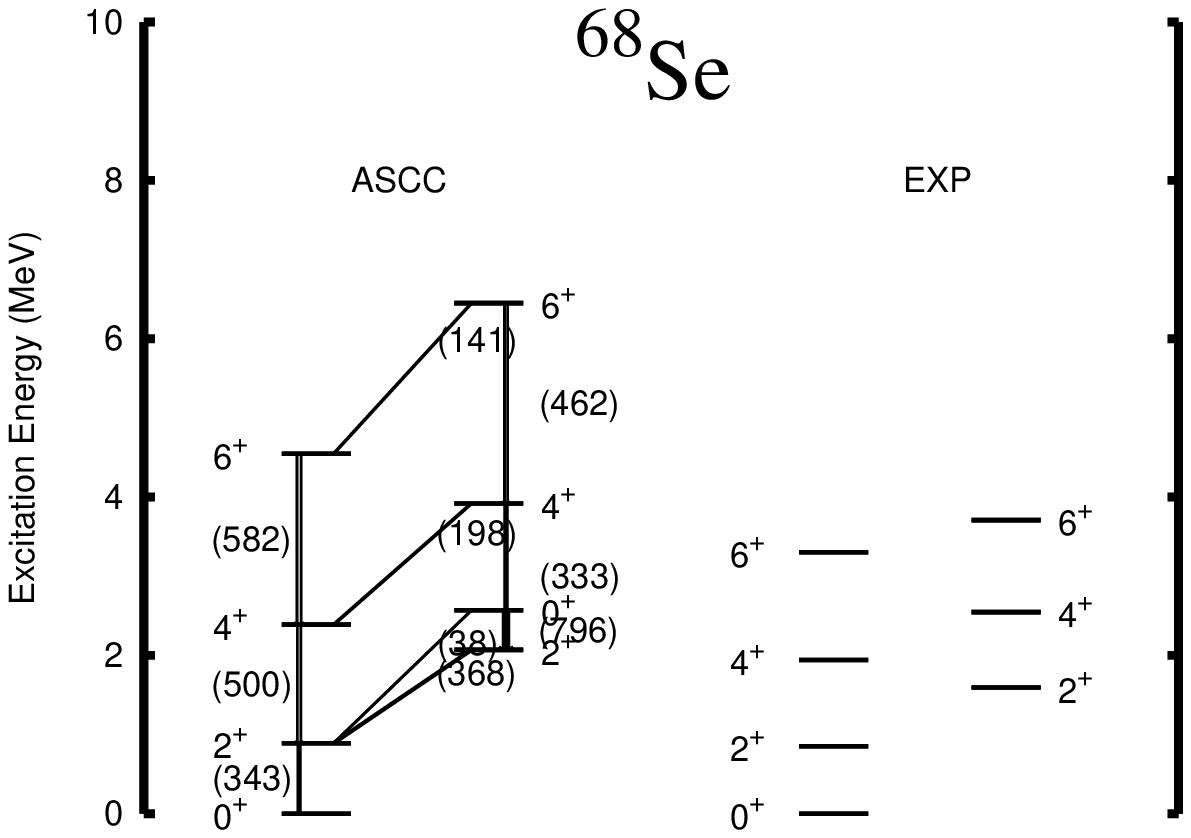}
&
\includegraphics[width=0.5\textwidth]{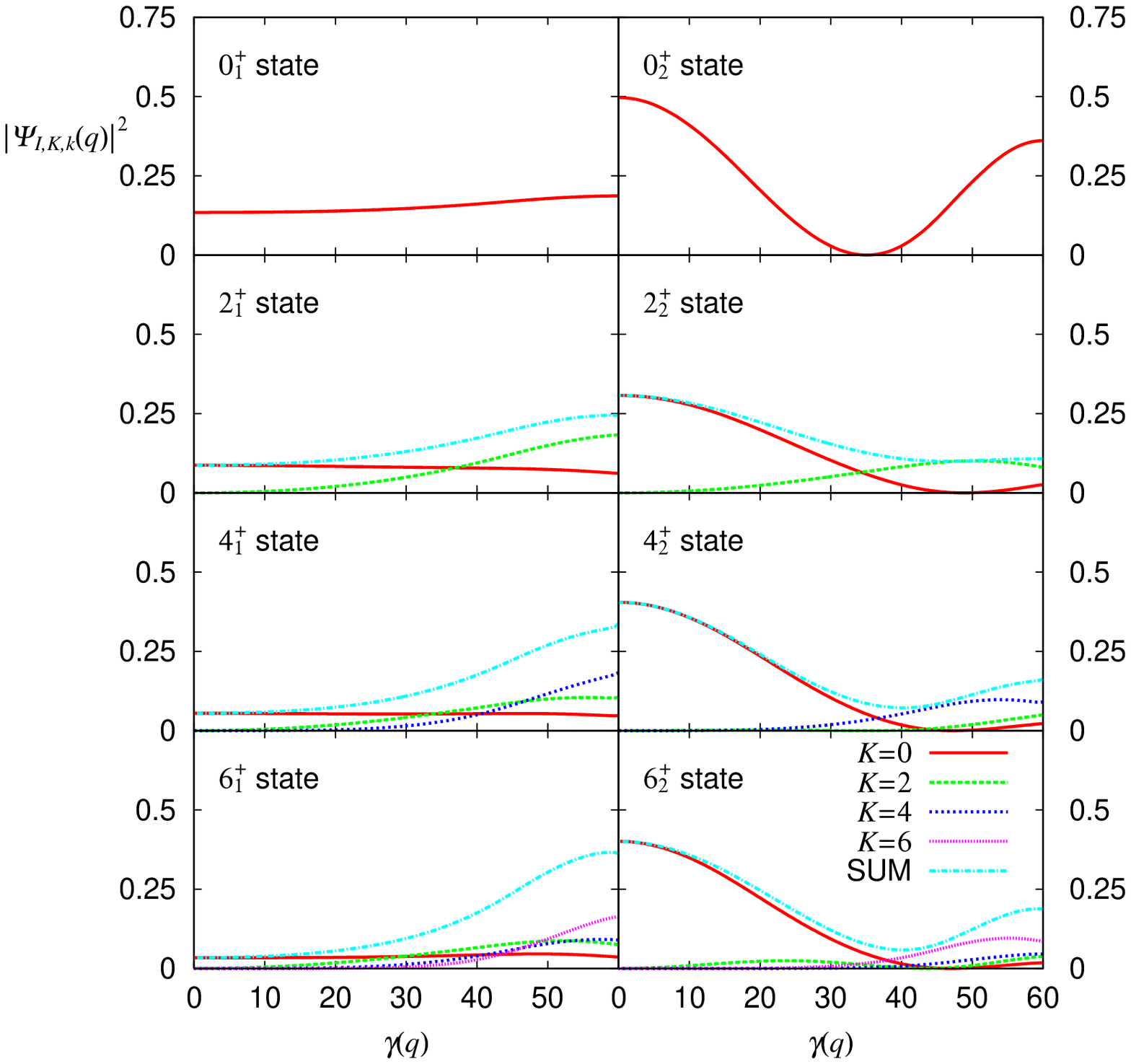}
\end{tabular}
\end{center}
\caption{
 Calculated spectra and $B(E2)\downarrow$ in units of e$^2$fm$^4$
(left) and the collective wave functions as functions of $\gamma$ (right).
The figures are taken from Ref.  \cite{HNMM09}
}
\label{fig: 68Se_spectra}
\end{figure}

\subsubsection{Discussion: Rotational localization}

In the left panel of Fig.~\ref{fig: 68Se_spectra}, excitation spectra
and $B(E2)$ values calculated for $^{68}$Se
are displayed together with experimental data.
The calculation yields two rotational bands, which
qualitatively agrees with experiment.
The calculated ground and excited bands
have oblate and prolate characters, respectively,
however, there is strong shape mixing, especially at low spins.
Note that the calculated $0_2^+$ state is located
above the $2_2^+$ state.
We find that this significant rise of the $0_2^+$ excitation energy
is due to the oblate-prolate shape mixing in the $0^+$ states,
which is much stronger than that of the $2^+$ states.
It would be very interesting to identify
the $0_2^+$ experimentally.\footnote{
In recent calculation with a (2+3)-dimensional collective Hamiltonian,
this state is predicted as the third $0^+$ state.
}

The strong shape mixing in the $0^+$ states can be confirmed
by examining the vibrational wave functions displayed
in the right panel of Fig.~\ref{fig: 68Se_spectra}.
Since the potential barrier between the two minima is only
400 keV high along the collective path,
this strong mixing is reasonable.
In fact,
the unusual behavior of the excited $0^+$ state suggests
an intermediate situation between the oblate-prolate shape coexistence
and the $\gamma$-unstable model by Wilet and Jean  \cite{WJ56}.

It is quite interesting to notice that the shape mixing becomes weak
as the angular momentum increases.
The collective wave functions of the
the $4^+$ and $6^+$ states tend to localize in the region near
either the oblate or the prolate shape.
Namely, it becomes more appropriate to characterize
the $4^+$ and $6^+$ states as oblate-like or prolate-like.
We have analyzed dynamical origin of this trend and found that
the rotational energy plays a crucial role in determining
localization of the collective wave function.
Therefore, this effect may be called
``rotational localization of collective wave function"
or ``rotational hindrance of shape mixing" \cite{HNMM09}.

\subsection{Further developments}

\subsubsection{Multi-dimensional collective submanifold}

\begin{wrapfigure}{r}{0.35\textwidth}
\includegraphics[width=0.35\textwidth]{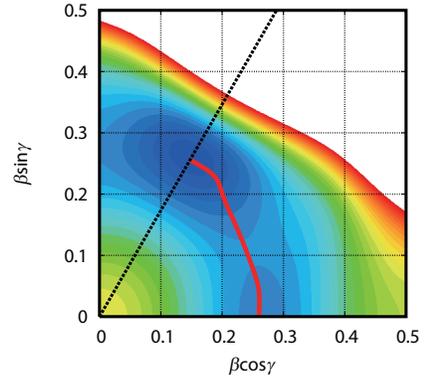}
\caption{\label{fig: 68Se_path}
The collective path for $^{68}$Se in the $(\beta,\gamma)$ plane.
The contour plot shows the result of the energy minimization with
the constraints on the mass quadrupole operators, $D_{20}$
and $D_{22}+D_{2-2}$.
}
\end{wrapfigure}

The 1D collective path for $^{68}$Se is obtained by following the
lowest eigensolution of the MFHE.
Figure \ref{fig: 68Se_path} shows an embedded collective path
in the $(\beta,\gamma)$ plane.
However, 
the frequency of the second lowest solution is only $1\sim 1.5$ MeV higher
than the lowest one (Fig.~\ref{fig: 68Se_ASCC}).
Thus, the extension from 1D to 2D may be important.
We often encounter similar situations which suggest importance
of the multi-dimensional collective submanifold.
Numerically, this is a challenging task, because we need to search for
self-consistent solutions of the MFHE in the 2D or higher-dimensional
hypersurface.
The technical developments for the multi-dimensional collective submanifold
is an important future subject.

In recent papers \cite{HSNMM10,SH11,HSYNMM11},
the collective Hamiltonian with the 2D shape degrees has been
constructed assuming the one-to-one correspondence between the
collective coordinates $(q^1,q^2)$ and $(\beta,\gamma)$,
combined with following approximations:
\begin{enumerate}
\item The collective submanifold is determined by the minimization
    with respect to the mass quadrupole operators, neglecting
    the self-consistency between Eqs. (\ref{CDFT}) and (\ref{MFHE}).
\item In the MFHE, the curvature terms are neglected.
\end{enumerate}
The eigensolutions of the MFHE are used to calculate the collective
mass parameters, $\bar{B}^{ij} = f^i_{,\alpha} f^j_{,\beta}
\widetilde{B}^{\alpha\beta}$.
The inclusion of the 2D shape degrees of freedom turns out to
be qualitatively consistent with the 1D calculation
in Fig.~\ref{fig: 68Se_spectra}, but further
improve the results in comparison with experiments \cite{HSNMM10}.

\subsubsection{Applications to modern energy functionals}

It is also highly desired to apply the method to realistic modern
energy functionals, such as Skyrme, Gogny, and covariant energy functionals.
This requires a developments of the code for MFHE without the
axial symmetry restriction.

Adopting the similar approximations mentioned above,
there are numerous recent works,
in the Skyrme energy functionals \cite{PQSL04,PR09}
the Gogny energy functionals \cite{LGD99,Del10},
and the covariant EDFs \cite{Nik09,Li09},
to construct the collective Bohr Hamiltonian with
the microscopic inputs.
However, the collective mass parameters are further approximated
by the Inglis-cranking formula.
Since this cranking approximation has a well-known defect leading
to a disagreement with experiments,
the phenomenological scaling to increase the collective mass by
roughly $30-40$ \%, is often adopted.
This comes from the fact that the cranking formula neglects the
time-odd mean fields \cite{RS80}.
Therefore, the replacement of the cranking mass by the MFHE mass
should solve this problem.

The MFHE with the Skyrme energy functional
has been applied to studies of Cr isotopes
with the above approximations and a restriction to the 1D shape degree
 \cite{YH11}.
In a very recent work \cite{HLNNV12},
a hybrid model, combining the covariant EDF
method with the MFHE in the P+Q model, has been investigated.
Namely, the potential energy surface is calculated by the
covariant EDF and the collective mass parameters
are estimated by the help of MFHE in the P+Q model.
The model shows a significant improvement in the low-lying spectra
in $\gamma$-soft nuclei of Xe and Ba isotopes.
These results suggest a promising future of the MFHE in the
energy functional approaches.

\section{Summary}
\label{sec: summary}
Intensive studies in density functional theories (DFT) in recent years
have produced numerous new results and new insights into nuclear structure.
It is beyond the scope of the present paper to review all of these
developments.
In this paper, we mainly focus our discussion on the basic concepts
of the nuclear DFT and applications of the time-dependent DFT (TDDFT).

The nucleus is a self-bound isolated system without an external
potential.
This produces a situation different from many-electron systems
with the external Coulomb potentials.
Therefore, we need modify the original arguments of the DFT, such as
the Hohenberg-Kohn theorem.
In Sec.~\ref{sec: basic_formalism},
we presented a justification based on the Hohenberg-Kohn theorem
modified for the wave packet.
It can be formulated with the Kohn-Sham scheme as well.
To incorporate effects of the nucleonic pair condensation,
the density functional should be generalized to include the
pair density (abnormal density) in addition to the normal density.
This can be done with the generalized scheme by Bogoliubov.
The TDDFT was also established,
in the same manner as the Hohenberg-Kohn, by the
one-to-one correspondence between the potential and density
depending on time.

Most applications of the TDDFT are performed in the perturbative
linear response regime.
Even in the linear regime, because of the
complexity of the EDFs of nuclei,
it is computationally very demanding to perform the numerical
calculations and requires significant works for its program coding.
There are several approaches to the linear-response calculations,
each of which has an advantage and disadvantage.
This has been discussed in details in Sec.~\ref{sec: perturbative_regime}.
As a feasible approach, we have proposed the finite
amplitude method (FAM).
The FAM only requires a minor modification of the existing program
of the stationary calculation for the nuclear ground-state properties.

The nucleus is also known to show many collective phenomena at low energies
which are not able to be described by the linear-response theory.
A typical example is given by the fission of heavy nuclei,
and by the shape phase transition and the shape coexistence.
For this purpose, we have presented a theory of a decoupled collective
space inside the large TDDFT phase space in Sec.~\ref{sec: LACM}.
In case that the collective motion of interest is slow relative to
the other intrinsic degrees of freedom,
self-consistent solutions of the moving-frame harmonic equation (MFHE)
with the constraint minimization of the EDF provide microscopic
determination of the collective variables and the collective Hamiltonian.

In Sec.~\ref{sec: GR} and Sec.~\ref{sec: application_LACM},
we presented selected results of recent studies
with the TDDFT on nuclear dynamics in the linear-response regime and
beyond the linear regime.
Properties of giant resonances in nuclei, especially those of the
giant dipole resonance (GDR), have been studied with
the Skyrme energy functionals.
Currently available EDFs are able to reproduce the GDR in heavy
nuclei, however, have a problem for that in light nuclei.
We also presented the applications of the MFHE to shape coexistence
phenomena in $^{68}$Se.
The calculation suggests a rotation-induced localization of the
collective wave functions, which produces the oblate-prolate
shape coexistence in this nucleus.

An interplay between theory and experiment has been and will be
providing deeper understanding of the nuclear quantum
many-body problem.
To develop a comprehensive predictive theory of the
nucleus to answer fundamental scientific questions,
still significant components are missing from our current
understanding.
Some of these missing parts can be addressed only by the
large-scale numerical simulations of nuclear systems.
The petascale- and of future exascle-computing platforms
is expected to pave the way to this goal.


\ack

I thank all my collaborators working together on subjects in this paper,
including
 P. Avogadro, G. Do Dang, S. Ebata, T. Inakura, N. Hinohara,
 M. Matsuo, K. Matsuyanagi, K. Sato,
 N. Walet,  K. Yabana, and K. Yoshida.
Part of the work was carried out under Grant-in-Aid for Scientific
Research(B) No. 21340073 and Innovative Areas No. 20105003.
The numerical calculations were performed in part on the
RIKEN Integrated Cluster of Clusters (RICC), and
on PACS-CS and T2K in University of Tsukuba.

\end{document}